\documentclass[%
article,
 amsmath,amssymb,
 aps,
]{revtex4-2}

\usepackage{graphicx}
\usepackage{dcolumn}
\usepackage{bm}
\usepackage{multirow}

\begin{document}

\preprint{APS/123-QED}

\title{SI-Traceable Temperature Calibration Based on Optical Lattice Clocks}

\author{Jin Cao}
\author{Benquan Lu}

\author{Xiaotong Lu}
\email{luxiaotong@ntsc.ac.cn}
\author{Hong Chang}
\email{changhong@ntsc.ac.cn}
\affiliation{National Time Service Center, Chinese Academy of Sciences, Xi'an 710600, China}
\affiliation{School of Astronomy and Space Science, University of Chinese Academy of Sciences, Beijing 100049, China}
\affiliation{Hefei National Laboratory, 230088 Hefei, China}

\date{\today}

\begin{abstract}
  We present a SI-traceable temperature calibration apparatus utilizing optical lattice clocks for precision metrology. The system employs a dual-blackbody radiation shield chamber with independent temperature control, enabling synchronous differential measurements of blackbody radiation (BBR)-induced frequency shifts in atomic ensembles. By correlating these shifts with chamber temperature, we propose absolute temperature determination traceable to the SI second through the optical clock frequency. Comprehensive uncertainty analysis demonstrates an absolute temperature uncertainty below 17 mK across the $200 \sim 350$ K range based on $^{87}$Sr optical lattice clock, representing an improvement of two orders of magnitude over current temperature measurements based on BBR-induced Rydberg state transitions. This advancement in primary thermometry offers significant improvements in precision, reproducibility, and versatility, with potential applications in metrology, fundamental physics, and industrial processes. 
\end{abstract}

\maketitle


\textit{Introduction.}—Among the seven fundamental physical quantities in the International System of Units (SI), temperature holds a pivotal role in preserving the coherence and integrity of the SI framework\cite{international2001international,thompson2008use}. The Kelvin scale, as the absolute standard for temperature, provides a universal reference that bridges diverse scientific domains—from thermodynamics and quantum mechanics to statistical mechanics—ensuring uniformity in one of nature's most fundamental parameters\cite{deutsch1991quantum,buvca2019non}.

The International Temperature Scale of 1990 (ITS-90), along with other conventional scales, relies on empirical definitions rooted in the phase transitions or radiative properties of materials such as water, hydrogen, and silver\cite{preston1990international,mangum1990guidelines,hill2014international}. Thus, these scales are highly sensitive to the physical states of substances, including pressure, purity, and chemical composition. Variations in the properties of these substances, which can occur under different experimental conditions, may lead to instability and inaccuracies in temperature calibration\cite{fellmuth2006determination,childs2000review}.

In 2018, the kelvin was redefined by the 26th General Conference on Weights and Measures through the assignment of the Boltzmann constant to \( k_B = 1.380649 \times 10^{-23} \)\,J/K, allowing for direct thermodynamic temperature measurements that are independent of material properties\cite{fischer2015progress,machin2018kelvin,rourke2024future}. Despite this advancement, practical temperature calibration in the range of hundreds of kelvins continues to rely on traditional methods, such as metal fixed-point cells and precision thermostatic baths, which are constrained by material-induced uncertainties. Recent efforts at NIST introduced a calibration-free approach utilizing BBR-induced transitions in Rydberg states of alkali-metal atoms, achieving SI-traceable thermometry\cite{ovsiannikov2011rydberg,schlossberger2025primary}. Nevertheless, challenges arising from stray electric fields and frequency measurement uncertainties led to an absolute uncertainty of 2 K at room temperature, highlighting the need for further improvements.

In this letter, we present an SI-traceable temperature calibration system based on optical lattice clocks. It employs a dual-BBR shield architecture with independent temperature control, where a reference chamber is stabilized at 4.2 K and a measurement chamber is regulated between $200 \sim 350$ K. A platinum resistance thermometer (PRT) in the measurement chamber enables temperature determination via a synchronous differential measurement protocol, linking atomic transition frequency shifts to temperature\cite{bothwell2022resolving,takamoto2011frequency}. Finite element simulations and Monte Carlo sampling indicate an absolute temperature uncertainty below 17 mK across the $200 \sim 350$ K range. This advancement in primary thermometry enables precise PRT calibration and, with a transportable optical lattice clock, facilitates versatile temperature calibration for diverse applications.

\textit{BBR Shift Calculation.}—The blackbody radiation frequency shift is the dominant shift in state-of-the-art $^{87}$Sr and $^{171}$Yb optical lattice clocks\cite{takamoto2005optical,zhang2022ytterbium,beloy2014atomic}.  
The BBR shift between the ground state \( g \) and excited state \( e \) is expressed as\cite{middelmann2012high}
\begin{equation}
  \begin{aligned}
  \Delta\nu_{e g}(T) =  - \frac{1}{{2h}}\Delta {\alpha _{\text{stat}}}{\left\langle {{E^2}} \right\rangle _T}[1 + \eta (T)] \\ = \Delta {\nu ^{(\text{stat})}}(T) + \Delta {\nu ^{(\text{dyn})}}(T),
  \end{aligned}
\end{equation}
where \(\Delta\alpha_{\text{stat}}\) denotes the static differential polarizability between the ground state \( g \) and the excited state \( e \), and \(\left\langle {{E^2}(T)} \right\rangle\) represents the mean-square electric field of the blackbody radiation. According to Planck's law, \(\left\langle {{E^2}(T)} \right\rangle  = {[8.319430(15)\,\text{V/cm}]}^2(T/{300\,\text{K}})^4\). 
The term $\eta(T)$ represents a minor dynamic correction factor associated with the frequency-dependent polarizability $\Delta\alpha(\omega)$. When multiplied by the mean square electric field, it constitutes the dynamic correction $\Delta \nu^{(\text{dyn})}(T)$ to the blackbody radiation shift.
For Sr clocks at room temperature, the fractional BBR shift is measured as \( -4.84172 \times 10^{-15} \), with an uncertainty of \( 7.3 \times 10^{-19} \)\cite{aeppli2024clock}.
High-precision measurements of the static polarizability $\Delta \alpha_{\text{stat}}$ have reduced the uncertainty contribution from $\Delta\nu^{(\text{stat})}$ to $1.4 \times 10^{-19}$ at $300\,\text{K}$\cite{lisdat2021blackbody}. As a result, the dominant source of uncertainty in the BBR shift is now attributed to the dynamic correction $\Delta \nu^{(\text{dyn})}(T)$.

Using the differential dynamic shift between the clock states $e$ and $g$, which is associated with the Einstein coefficient $A$, the dynamic frequency shift $\Delta \nu^{(\text{dyn})}(T)$ can be calculated. For an arbitrary state $i$, frequency shift is given by\cite{lisdat2021blackbody}
\begin{equation}
\delta \nu _i^{(\text{dyn})}(T) =  - \frac{1}{{4\pi }}{(\frac{{{k_B}T}}{h})^3}\sum\limits_k {\frac{{2{J_k} + 1}}{{2{J_i} + 1}}\frac{{{A_{ki}}}}{{\nu _{ik}^3}}G(\frac{{h{\nu _{ik}}}}{{{k_B}T}})},
\end{equation}
where \( G \) is an integral function, and \(\Delta \nu^{(\text{dyn})}(T) = \delta \nu_e^{(\text{dyn})}(T) - \delta \nu_g^{(\text{dyn})}(T)\). For Sr, the $5s4d~^3\!D_1 \to 5s5p~^3\!P_0$ transition contributed 98.2\% of the dynamic correction to the $^3\!P_0$ state\cite{safronova2013blackbody}. Aeppli \textit{et al.} measured the lifetime of the $^3D_1$ state as $\tau = 2.156(5)\,\mu\text{s}$\cite{aeppli2024clock}. By combining this result with other the most recent and lowest-uncertainty experimental measurements, including the 813 nm and 390 nm magic wavelengths and the tune-out wavelength, we employed a self-consistent approach to optimize seven Einstein $A$ coefficients (see Appendix) for the calculation of $\Delta \nu^{(\text{dyn})}(T)$, and the computational results are available in \cite{jin2025sr}. Monte Carlo sampling was employed to evaluate the dynamic BBR shift uncertainty, yielding \(\Delta \nu^{(\text{dyn})}(300 \, \text{K}) = -152.89(29) \, \text{mHz}\) at 300 K. Our result falls within the uncertainty range of the previous calculation \(-153.06(33)\)~mHz from Ref.\cite{aeppli2024clock}.
\begin{figure}[htbp]
  \centering
  \includegraphics[width = 8cm]{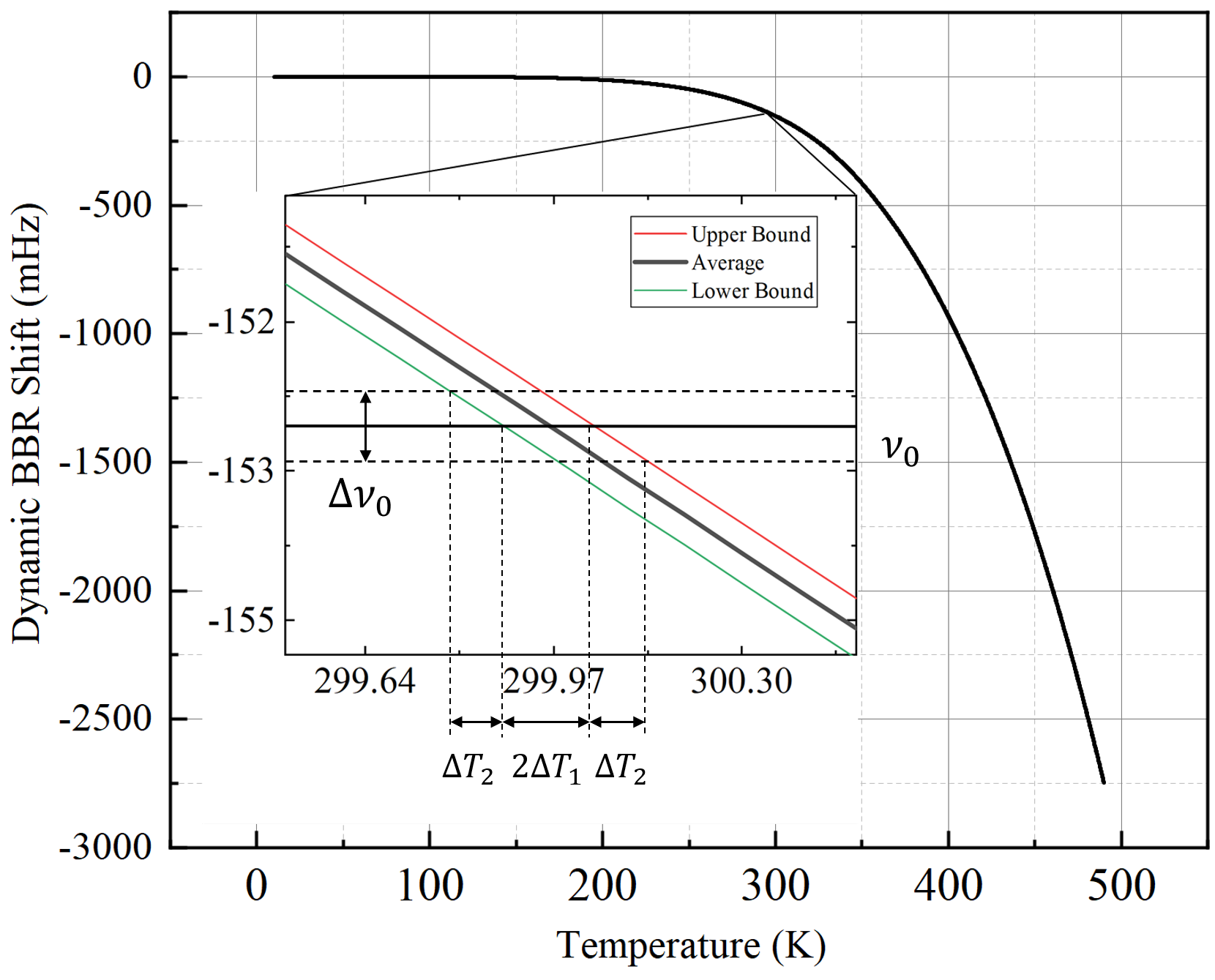}
  \caption{Dynamic blackbody radiation frequency shift for the Sr optical lattice clock at different temperatures. The red and green lines represent the uncertainty ranges of the frequency shift calculated using Monte Carlo sampling. \( \Delta T_1 \) and \( \Delta T_2 \) denote the absolute uncertainties in temperature calibration when measuring the frequency. }
  \label{fig1}

\end{figure}

Combining static and dynamic contributions, we determined the total BBR shift for \( ^{87} \text{Sr} \) optical lattice clocks across $4 \sim 490$ K with \( 10^{-19} \) level uncertainties. The high accuracy of the BBR shift calculation enables precise inference of external radiation source temperatures, provided other shift contributions are well-controlled.

\begin{figure*}[htbp]
  \centering
  \includegraphics[width = 18cm]{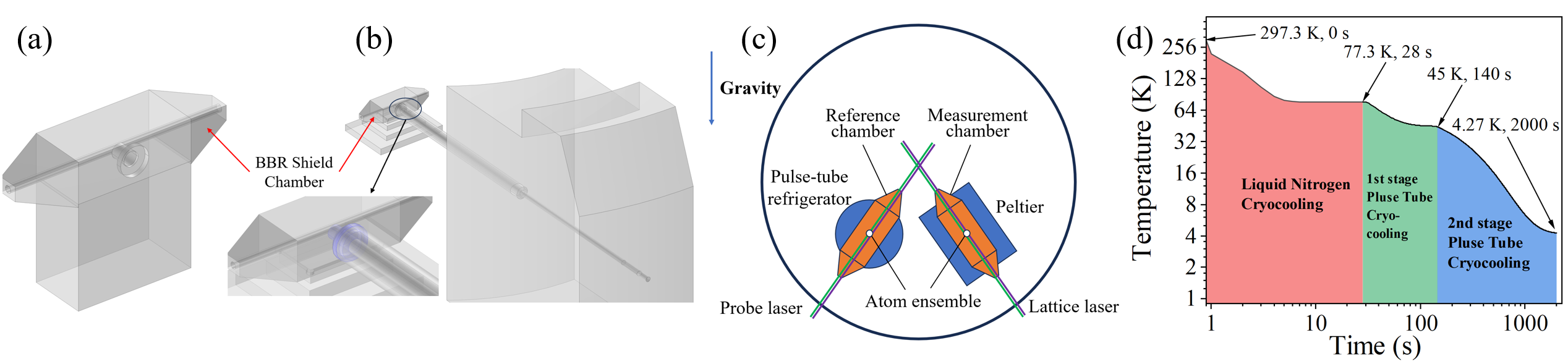}
  \caption{(a) Simulation model of the reference chamber; (b) Simulation model of the measurement chamber; (c) Schematic diagram of the experimental design; (d) Temperature variation over time at different cooling stages for the reference chamber.}
  \label{fig2}

\end{figure*}

\textit{Experimental Design and Simulation.}—Accurate inference of environmental temperature from the frequency shift of an optical lattice clock requires precise control of radiation sources and isolation of the blackbody radiation shift from other frequency shifts. In a typical vacuum chamber housing a one-dimensional Sr optical lattice clock, multiple thermal sources contribute to the overall BBR shift\cite{ushijima2015cryogenic}. These include the Sr oven, the vacuum maintenance device, and external environmental radiation leaking through the observation window. The total BBR shift is given by
\begin{equation}
    \Delta \nu_{\text{all}} = \sum_i \frac{\Omega_i}{4\pi} \Delta \nu_i(T_i),
\end{equation}
where \(\Delta \nu_i(T_i)\) is the BBR shift from the \(i\)-th source at temperature \(T_i\), and \(\Omega_i\) is the solid angle subtended by the source. This summation makes it challenging to disentangle individual contributions.

To address this, we propose to implement a blackbody radiation shielding chamber (Fig.~2). Quantitative analysis of the environmental radiation exposure reveals a power of $3.006 \times 10^{-4}$\,W for unshielded atoms, compared to merely $1.66 \times 10^{-7}$\,W at the chamber center. This three-order-of-magnitude reduction demonstrates that the chamber attenuates $3.004 \times 10^{-4}$\,W of environmental radiation, achieving a shielding efficiency of 99.88\%. Through real-time monitoring and active temperature stabilization, we expect the chamber to maintain a short-term temperature stability below 1 mK. The inner walls (atomic channel) are coated with high-emissivity carbon nanotube layers. Owing to the unique forest-like microscopic structure of carbon nanotubes, their emissivity shows angular dependence\cite{mizuno2009black} We have computed the polar-angle dependence of the emissivity at the atomic cloud surface relative to the atomic channel axis, with full details provided in the Appendix. Using the angular-averaged emissivity as the effective value for the atomic channel, and taking the difference between the maximum and minimum values as the uncertainty, we obtain an effective emissivity uncertainty of 0.005. The chamber's thermal is monitored by a PRT at the center of the chamber. To further extract the BBR shift, we implemented a dual-chamber configuration. The reference chamber is cooled to 4.2\,K using a pulse-tube refrigerator, while the measurement chamber is temperature-adjustable ($200 \sim 350$ K) via a Peltier device. 

To stabilize the reference cavity temperature at 4.2 K, we propose a three-stage cooling protocol, with performance validated through finite-element simulations (Fig. 2d). The designed cooling sequence would begin with liquid-nitrogen evaporation cooling to 77 K, followed by activation of the 4 K pulse-tube cryocooler's first stage to reach 45 K. Upon reaching thermal equilibrium, the second cryocooler stage would be engaged to attain the target 4.2 K. Our design maintains the cooling head under vacuum to minimize thermal loads, with thermal connections limited to carefully insulated liquid-nitrogen and helium pipelines. The first cryocooler stage could provide 200 W of cooling power at 77 K (sufficient for cooling to 45 K), while the second stage offers 10 W. Simulation results indicate that by optimizing the thermal contact geometry and minimizing parasitic heat leaks, 4.2 K operation should be achievable. Projected temperature stabilization would be maintained through real-time monitoring and dynamic adjustment of the cryocooler power.

Our theoretically designed apparatus measures the atomic channel surface temperature within the measurement chamber to calibrate a platinum resistance thermometer (PRT). For this purpose, we incorporate an access port on the side of the chamber, terminating adjacent to the atomic channel. A PRT for calibration is positioned at the port terminus, maintaining robust thermal contact with the chamber wall via silicone grease to ensure temperature uniformity between the atomic channel surface and the PRT. To enable thermometer insertion and removal, we implement a quartz tube assembly enclosing the chamber port, connected to a vacuum feedthrough. This configuration permits the PRT to be inserted through the feedthrough and guided by the tube to its measurement position. The quartz tube preserves vacuum integrity while providing access. Finite element simulations verify that the thermally insulating materials in our design effectively mitigate thermal conduction from the vacuum chamber, maintaining a temperature gradient of less than 0.3 mK between the atomic channel surface and the PRT.

For BBR shift measurements, we propose to laser-cool the first Sr ensemble via two-stage cooling at the MOT (magneto-optical trap) cavity center and transport it into the reference chamber using a moving optical lattice\cite{schmid2006long,zheng2022differential}. The second ensemble, prepared identically, is transferred to the measurement chamber. Both ensembles undergo synchronous clock interrogation using two laser beams phase-locked to the same ultra-stable cavity. After interrogation, the atomic ensemble in the reference chamber are transported back to the MOT region for excitation fraction measurement. Subsequently, the excitation fraction of atomic ensemble in the measurement chamber is determined using the identical detection protocol.

The atomic channel walls in the reference chamber contribute merely 0.23\% of the total radiation intensity ($1.1\times10^{-3}$~W/m$^2$ of $0.48$~W/m$^2$ total), inducing a frequency shift below $10^{-7}$~Hz. This demonstrates that the blackbody radiation shift from the reference chamber itself is negligible. Consequently, the dominant contribution to the BBR shift experienced by atoms in the reference chamber arises from residual environmental radiation.

Figure~3(b) reveals that the radiation intensity difference between measurement and reference chambers at the atomic ensemble surfaces matches (within $1.1 \times 10^{-3}$~W/m$^2$) the isolated measurement chamber intensity. This agreement demonstrates that synchronous differential measurements effectively cancel environmental radiation frequency shifts. To assess environmental sensitivity, we model the vacuum chamber with a $350$~mK thermal gradient\cite{hobson2020strontium}. The analysis reveals maximum radiation intensity variations of $2.29\times10^{-3}$~W/m$^2$, producing a differential fractional frequency shift below $1.22\times10^{-20}$. These results demonstrate the measurement's intrinsic robustness against environmental thermal fluctuations at the $10^{-20}$ level.

Using the BBR shift-temperature relationship, the measurement chamber's temperature is determined with high accuracy. This quantum-based measurement serves as an absolute temperature standard for calibration.
\begin{figure}[htbp]
  \centering
  \includegraphics[width=13cm]{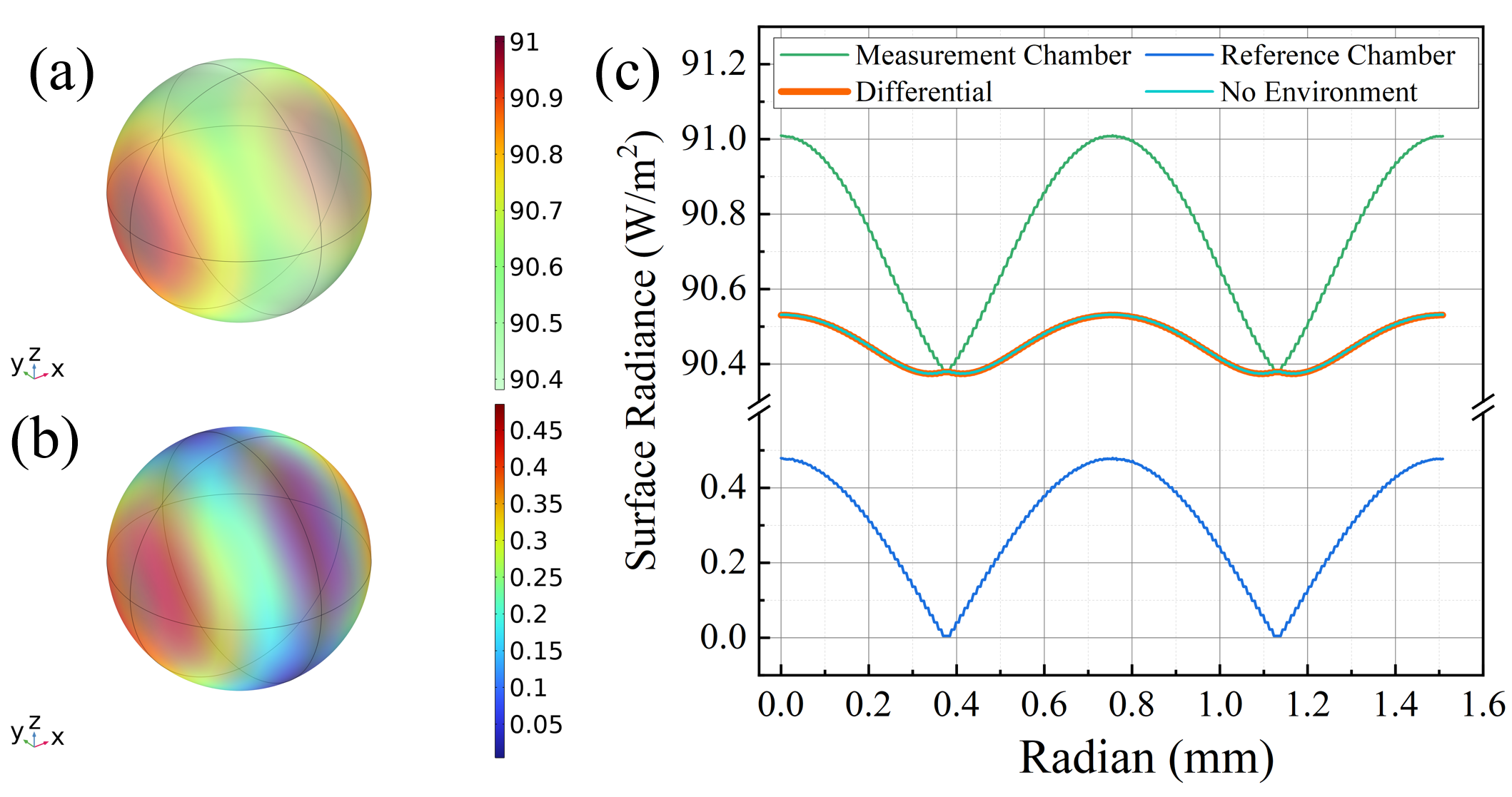}
  \caption{(a) Surface radiative intensity distribution in the reference chamber; (b) Surface radiative intensity distribution in the measurement chamber; (c) Radiation intensity on the atomic surface as a function of radians. Blue and green lines represent the reference and measurement chambers, respectively. The orange line indicates the difference, and the cyan line represents the measurement chamber under zero environmental radiation.}
  \label{fig3}
\end{figure}

\textit{Uncertainty Analysis}—The uncertainties associated with temperature calibration based on multiple atomic clock synchronous differential measurements are analyzed and compared to those of existing temperature calibration systems. The uncertainty in temperature calibration using synchronous differential measurements is primarily attributed to two sources: (1) the uncertainty in evaluating the atomic BBR shift and (2) the uncertainty arising from non-uniform temperature distribution within the shielding chamber.

Monte Carlo sampling is employed to evaluate the uncertainty in the BBR shift for the Sr atomic system. Figure~1 shows the BBR shift as a function of temperature (black line), computed using optimized Einstein coefficients \(A\). The red and green lines represent the upper and lower bounds of the BBR shift uncertainty derived from Monte Carlo sampling. From the result of a single synchronous differential measurement, \(\nu_0\), the temperature of the measurement chamber is inferred using the BBR shift-temperature relationship. This introduces a temperature uncertainty \(\Delta T_1\) due to the BBR shift evaluation uncertainty. Calculations of \(\Delta T_1\) for both Sr and Yb atomic systems at various temperatures are summarized in Table~1.

In practice, the measurement frequency \(\nu_0\) exhibits inherent uncertainty due to the stability limitations of the optical lattice clock, introducing an additional temperature uncertainty $\Delta T_{2}$. For differential frequency measurements, state-of-the-art Sr optical lattice clocks demonstrate remarkable stability characterized by \(4.4 \times 10^{-18}/\sqrt{\tau}\), where $\tau$ represents the integration time in seconds\cite{bothwell2022resolving}. This stability enables achieving fractional frequency uncertainties below $1 \times 10^{-19}$ with 2000\,s of averaging time. However, the ultimate precision is constrained by systematic effects, including the lattice AC Stark shift, second-order Zeeman shift, density shift, background gas collision shift, and gravitational redshift\cite{kim2023evaluation,denker2018geodetic}.

Through careful experimental design, several systematic shifts can be effectively mitigated. By precisely aligning the measurement and reference chambers at identical horizontal positions and equal distances from the magneto-optical trap center, we eliminate contributions from both the second-order Zeeman shift and gravitational redshift. Although the reference and measurement cavities reside in the same vacuum environment, cryogenic pumping of the $4$~K chamber can introduce residual pressure differences between them, necessitating independent characterization of background gas collision shifts for each ensemble\cite{pagano2018cryogenic}. For Sr optical lattice clocks, this shift is suppressed below $10^{-19}$ at pressures under $7\times10^{-9}$~Pa\cite{rindal2019effect}, whereas Yb clocks require pressures below $3\times10^{-8}$~Pa\cite{mcgrew2018atomic}. We will precisely quantify the differential collisional shift by systematically varying the trap lifetime between the two lattice sites. By modulating the getter pump temperature or the ion pump speed, we could control the trap lifetime limited by residual pressure, enabling direct experimental determination of this shift contribution at the $10^{-19}$ level.

Regarding the density shift, although our optical lattice is not oriented vertically but rather at an angle with respect to the direction of gravity, our calculations show that at a tilt angle of $30^\circ$, the magic trap depth required to cancel the on-site $p$-wave and off-site $s$-wave interactions is $12.8\,E_r$, which still lies within the shallow lattice regime (see Appendix for detailed calculation). Furthermore, even in the presence of small residual collisional shifts, we can accurately characterize and subtract them at the level below $1 \times 10^{-19}$ using a differential measurement scheme.

The dominant residual uncertainty arises from the lattice AC Stark shift, which typically contributes approximately $5 \times 10^{-19}$ to $\nu_{0}$ even in the shallow lattice configuration\cite{kim2023evaluation}. Crucially, our synchronous differential measurement technique enables precise determination and cancellation of the differential AC Stark shift (see Appendix for details). Accounting for practical considerations such as lattice laser frequency and power fluctuations, we conservatively estimate the AC Stark shift-induced uncertainty to be $1 \times 10^{-19}$. We have conducted a comprehensive analysis of the temperature calibration uncertainty arising from frequency measurement inaccuracies, examining its dependence on both integration time and calibration temperature, with the results summarized in Fig.~4(b).

The temperature at the surface of the atomic channel is derived from measurements of the BBR shift. Due to geometric constraints, the PRT employed for calibration cannot be directly placed on the atomic channel surface. To minimize this limitation, the PRT is placed in close proximity to the atomic channel by insertion through a drilled hole. Caused by heating from absorption of lattice laser radiation during its passage through the atomic channel and the thermal conductivity of the metal, a temperature gradient \(\Delta T_3\) arises between the atomic channel surface and the PRT. We characterize \(\Delta T_3\) as the temperature calibration uncertainty arising from thermal conduction within the measurement chamber.
To quantify \(\Delta T_3\), finite element simulations at various temperatures within the measurement cavity was performed. The simulation details are provided in the Appendix, and the results are presented in Fig.~4(c). Given the high thermal conductivity of copper, the material constituting the cavity, the finite element simulation results demonstrate that \(\Delta T_3\) remains below 0.5 mK at 200s. 

The angular dependence of the carbon nanotube emissivity contributes an emissivity uncertainty, which we propagate to the temperature calibration uncertainty $\Delta T_{emission}$. As detailed in the Appendix, the resulting emission-induced temperature uncertainty reaches a maximum of $\Delta T_{emission} = 0.6 $~mK across the measured cavity temperature range.
The overall temperature calibration uncertainty is expressed as 
\begin{equation}
\Delta T_{\text{total}} = \sqrt{(\Delta T_1)^2 + (\Delta T_2)^2 + (\Delta T_3)^2 + (\Delta T_{emission})^2}.
\end{equation}

\begin{figure}[h]
  \centering
  \includegraphics[width = 13cm]{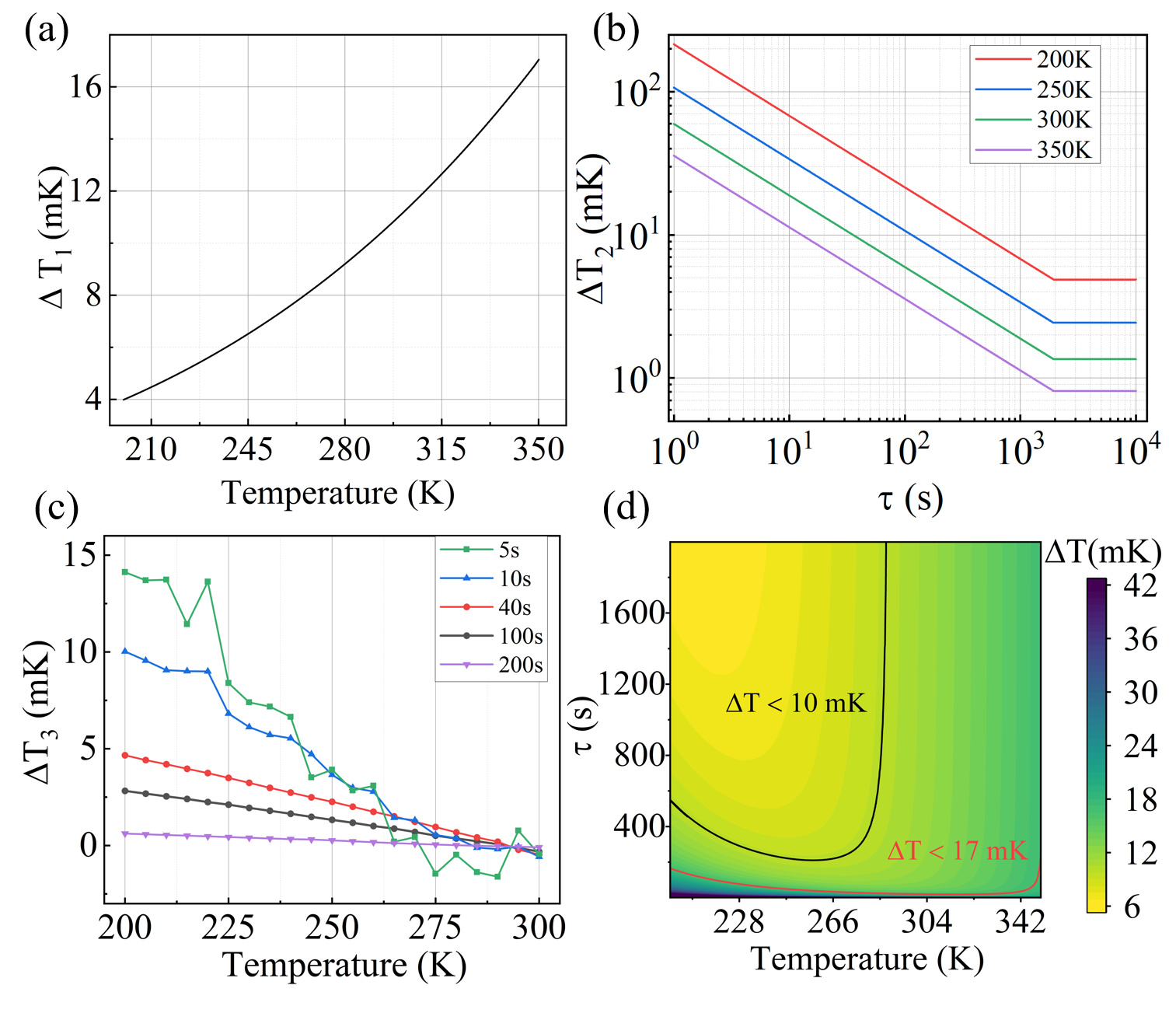}
  \caption{(a) \(\Delta T_1\) at different calibration temperatures. 
  (b) \(\Delta T_2\) as a function of measurement time for different calibration temperatures. 
  (c) \(\Delta T_3\) at different calibration temperatures and thermal conduction times. 
  (d) Total \(\Delta T\) as a function of measurement time and calibration temperature. 
  The region enclosed by the red line indicates where \(\Delta T < 17\)~mK. }
  \label{fig4}
\end{figure}

We calculated \(\Delta T_{\text{total}}\) for different calibration temperatures and measurement times, and the results are shown in Fig.~\ref{fig4}(d). The region enclosed by the red line indicates \(\Delta T_{\text{total}} < 17\)~mK, demonstrating that our setup propose a temperature calibration with an uncertainty better than 17~mK over the range of $200 \sim 350$ K, which is on par with the calibration uncertainty of the First-Class Platinum Resistance Thermometer Standard.

We further computed and analyzed the temperature calibration uncertainties for various neutral atom optical lattice clocks serving as temperature standards. The results are summarized in Table~1. Atoms such as magnesium and cadmium, which exhibit low sensitivity to blackbody radiation, are excellent candidates for atomic clock references. Their insensitivity to BBR, while advantageous for clock stability, simultaneously leads to larger uncertainties in temperature calibration when employed as temperature standards.
The Yb atomic system shares many similarities with the Sr system. Due to the reduced dynamic contribution to the BBR shift associated with the more UV-like transition in Yb, the evaluation of the BBR shift can more readily propose an uncertainty at the \(10^{-19}\) level. However, the dynamic BBR shift uncertainty for Yb is approximately three times larger than that of Sr, leading to a higher temperature calibration uncertainty\cite{heo2022evaluation}.

Although fixed-point methods achieve a calibration uncertainty of less than 5~mK, they are limited to calibrating platinum resistance thermometers at specific temperature points, such as the triple point of mercury (234.3156~K) or the triple point of water (273.16~K). Temperatures at other points must be interpolated. In contrast, our method enables continuous calibration across the $200 \sim 350$ K range. This capability provides significant advantages in versatility and precision. Furthermore, since the temperature calibration result depends solely on frequency measurements, it exhibits excellent reproducibility. When implemented in a portable design, an Sr optical lattice clock-based temperature calibration system can propose transportability comparable to standard platinum resistance thermometer devices.
\begin{table}[h]
  \centering
  \caption{Temperature calibration uncertainties for various neutral atom optical lattice clock systems.}
  \begin{ruledtabular}
  \begin{tabular}{lccccr}
  Item                           & $\Delta T_1 (mk)$ & $\Delta T_2 (mK)$ & $\Delta T_{total}(mK)$ & Ref. \\ \hline
  Sr(200K)                  & 3.98       & 4.88       & 6.34           & This work   \\
  Sr(250K)                    & 6.85      & 2.43        & 7.31          & This work   \\
  Sr(300K)                 & 11.07        & 1.35        & 11.17        & This work   \\
  Sr(350K)                    & 16.84      & 0.81        & 16.87        & This work   \\
  Yb(200K)                 & 11.81      & 57.53      & 58.73          & \cite{jin2023multiple}   \\
  Yb(300K)                  & 37.84      & 16.79      & 41.4          & \cite{jin2023multiple}   \\
  Mg(297K)                    & 3000             &      -        &      -      & \cite{wu2020magic}   \\
  Cd(300K)                   & 2032.98    &    -          &       -           & \cite{yamaguchi2019narrow}   \\
  Rydberg(300K)                  &              &              &  2000                & \cite{schlossberger2025primary}   \\ 
  First-Class              &       -       &    -          & 4-18           & \cite{NIM_CMC_2025}   \\
  PRT Standard             &              &              &                  &           \\
  Fixed-Point              &        -      &    -         & 0.5-5          & \cite{NIM_CMC_2025}   \\ 
  \end{tabular}
\end{ruledtabular}
  \label{method}
\end{table}

\textit{Conclusion.}—By converting temperature measurements into frequency measurements, we present a SI-traceable temperature calibration apparatus based on optical lattice clocks, achieving an absolute temperature uncertainty of less than 17 mK over the range of $200 \sim 350$ K. Utilizing a dual-chamber architecture combined with synchronous differential measurements of BBR-induced frequency shifts, we establish a direct connection between atomic transition frequencies and temperature, circumventing the reliance on material-based calibration methods. This approach not only propose calibration accuracy comparable to that of a First-Class PRT standard, but also enables continuous calibration across a broad temperature range. By integrating optical lattice clock technology with advanced BBR shielding and thermal control systems, we demonstrate a robust platform for high-precision thermometry. This system holds significant potential for applications in metrology, fundamental physics, and industrial processes, especially based on a transportable optical lattice clock. The currently calibration precision is mainly limited by the uncertainty of the BBR dynamic correction coefficient. The conceptual setup proposed in this work can be used to expediently and accurately determine the differential polarizability at specify wavelength\cite{lu2024determining,huntemann2016single}, which will improve the precision of BBR dynamic correction coefficient. Searching other clock transitions with higher sensitivity on the BBR shift could also further improve the precision of our method.

\textit{Ackonwledgement.}—This work is supported by the Innovation Program for Quantum Science and Technology (Grant No. 2021ZD0300902), the Strategic Priority Research Program of the Chinese Academy of Sciences (Grant No. XDB35010202), the Operation and Maintenance of Major Scientific and Technological Infrastructure of the Chinese Academy of Sciences (Grant No. 2024000014) and the National Natural Science Foundation of China (Grant No. 12203057).

\section{Appendix}

\subsection{ Lattice Light Shift and Density Shift}
\subsubsection{lattice light shift}
For the Sr optical lattice clock, the first-order Zeeman shift is described by the relation
\begin{equation}
\Delta\nu_\pi = -\delta g \, m_F \, \mu_0 B / h,
\end{equation}
where \(\delta g\) represents the differential \(g\)-factor, \(\mu_0\) is the Bohr magneton, and the subscript \(\pi\) denotes non-\(m_F\) changing transitions. To mitigate the impact of the first-order Zeeman shift, Sr atoms are typically polarized to the \((^3P_0, m_F=\pm9/2)\) states. As illustrated in Fig.~5, the transition frequencies \(f_1\) and \(f_2\) corresponding to \((^3P_0, m_F = +9/2 \rightarrow {^1S_0}, m_F=+9/2)\) and \((^3P_0, m_F = -9/2 \rightarrow {^1S_0}, m_F=-9/2)\) are measured alternately. By averaging these frequencies, the transition frequency \(f_0\) free from the first-order Zeeman shift is obtained. This procedure involving the measurement of \(f_1\) and \(f_2\) to derive \(f_0\) constitutes one measurement cycle.
\begin{figure}[h]
    \centering
    \includegraphics[width=12cm]{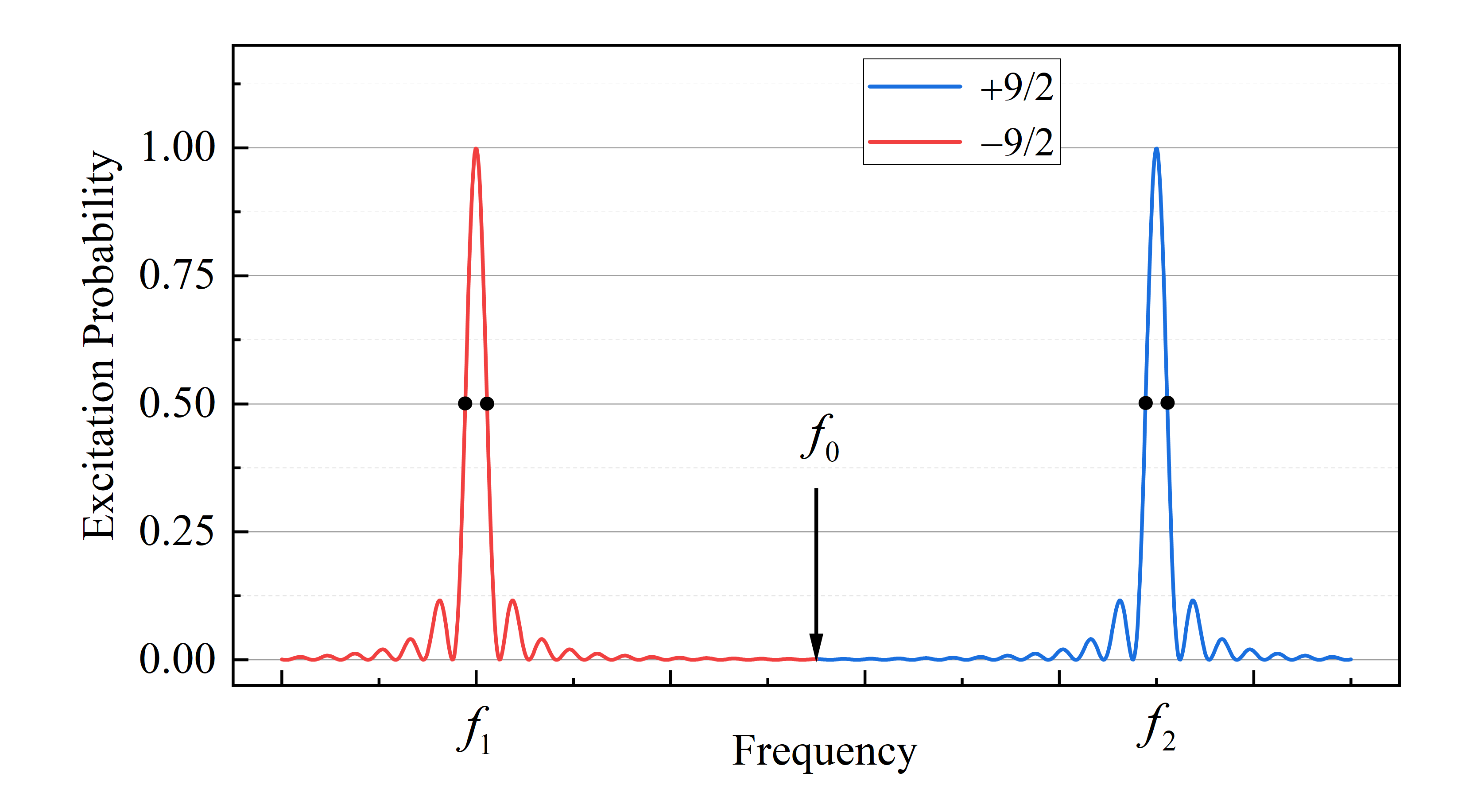}
    \caption{Clock operation utilizing stretched nuclear spin states for first-order Zeeman insensitivity. The center frequency $f_0$ is determined by averaging two independent locks, eliminating the linear Zeeman shift.}
    \label{fig10}
\end{figure}

In our theoretically proposed apparatus, the atoms in the measurement cavity and the reference cavity are trapped in two separate optical lattices, effectively functioning as two independent optical clocks, which we refer to as clock 1 and clock 2. As depicted in Fig.~1, the trap depths of these optical lattice clocks are maintained identically during operation. At a specific trap depth \( u_1 \), the frequency difference \( \Delta f_1 \) between the two clocks is determined using synchronous differential measurement techniques. The trap depth is then adjusted to \( u_2 \), and the measurement is repeated to obtain \( \Delta f_2 \). Given that the temperatures of BBR shielding cavities remain constant and the atoms in both cavities experience identical magnetic fields and vacuum conditions, the contributions of the BBR shift, collision shift, and second-order Zeeman shift to the difference \( \delta f = \Delta f_1 - \Delta f_2 \) are negligible. Consequently, \( \delta f \) is predominantly influenced by the lattice light shift and the density shift. With the density shift evaluable to an uncertainty of \( 9 \times 10^{-20} \) through measurements of trap depth and atomic ensemble density, the uncertainty of \( \delta f \) is primarily governed by the lattice light shift\cite{aeppli2024clock}. The lattice light shift is expressed as\cite{kim2023evaluation}:
\begin{equation}
h\Delta {f_{LS}}(u,{\delta _L},{n_z}) \approx \left(\frac{{\partial {\alpha ^{E1}}}}{{\partial f}}{\delta _L} - {\alpha ^{qm}}\right)\left({n_z} + \frac{1}{2}\right){u^{1/2}} - \left[\frac{{\partial {\alpha ^{E1}}}}{{\partial f}}{\delta _L} + \frac{3}{2}\beta \left(n_z^2 + {n_z} + \frac{1}{2}\right)\right]u  + 2\beta \left({n_z} + \frac{1}{2}\right){u^{3/2}} - \beta {u^2},
\end{equation}
where \(\delta_L\) denotes the lattice light detuning, and \(n_z\) is the axial state quantum number. For the initial measurement, the lattice light detunings for clock 1 and clock 2 are set to \(\delta_{L1}\) and \(\delta_{L2}\), respectively, with the trap depth fixed at \(u_1\) and the state \(n_z = 0\) selected. Synchronous differential measurements yield:
\begin{equation}
h\Delta f_1 = \frac{1}{2} \frac{\partial \alpha^{E1}}{\partial f} (\delta_{L2} - \delta_{L1}) u_1^{1/2} - \frac{\partial \alpha^{E1}}{\partial f} (\delta_{L2} - \delta_{L1}) u_1.
\end{equation}
Subsequently, adjusting the trap depth to \( u_2 \) and repeating the measurement provides:
\begin{equation}
h\Delta f_d = h(\Delta f_2 - \Delta f_1) = \frac{1}{2} \frac{\partial \alpha^{E1}}{\partial f} (\delta_{L2} - \delta_{L1}) (u_1^{1/2} - u_2^{1/2}) - \frac{\partial \alpha^{E1}}{\partial f} (\delta_{L2} - \delta_{L1}) (u_1 - u_2).
\end{equation}

As shown in Fig.~6(a), following these measurements, \( \delta_{L2} \) is altered, and the trap depth is reset to \( u_1 \). This cycle is repeated to accumulate sufficient data \( (\delta f_d, \delta_{L2}) \), which is then fitted to determine \( \frac{\partial \alpha^{E1}}{\partial f} \). Simulated experimental values are fitted, as illustrated in Fig.~6(b), and the intersection of the fitted curve with the \( x \)-axis identifies the lattice light detuning that nullifies \( \delta f \). At this detuning, the uncertainty in the light shift evaluation is dominated by the lattice light frequency (or wavelength) measurement uncertainty. With current technology, this frequency can be measured with an uncertainty on the order of mHz, corresponding to a relative frequency uncertainties below \( 1 \times 10^{-19} \). Conservatively, we adopt \( 1 \times 10^{-19} \) as the fractional frequency uncertainties of lattice light shift.
\begin{figure}[h]
    \centering
    \includegraphics[width=17cm]{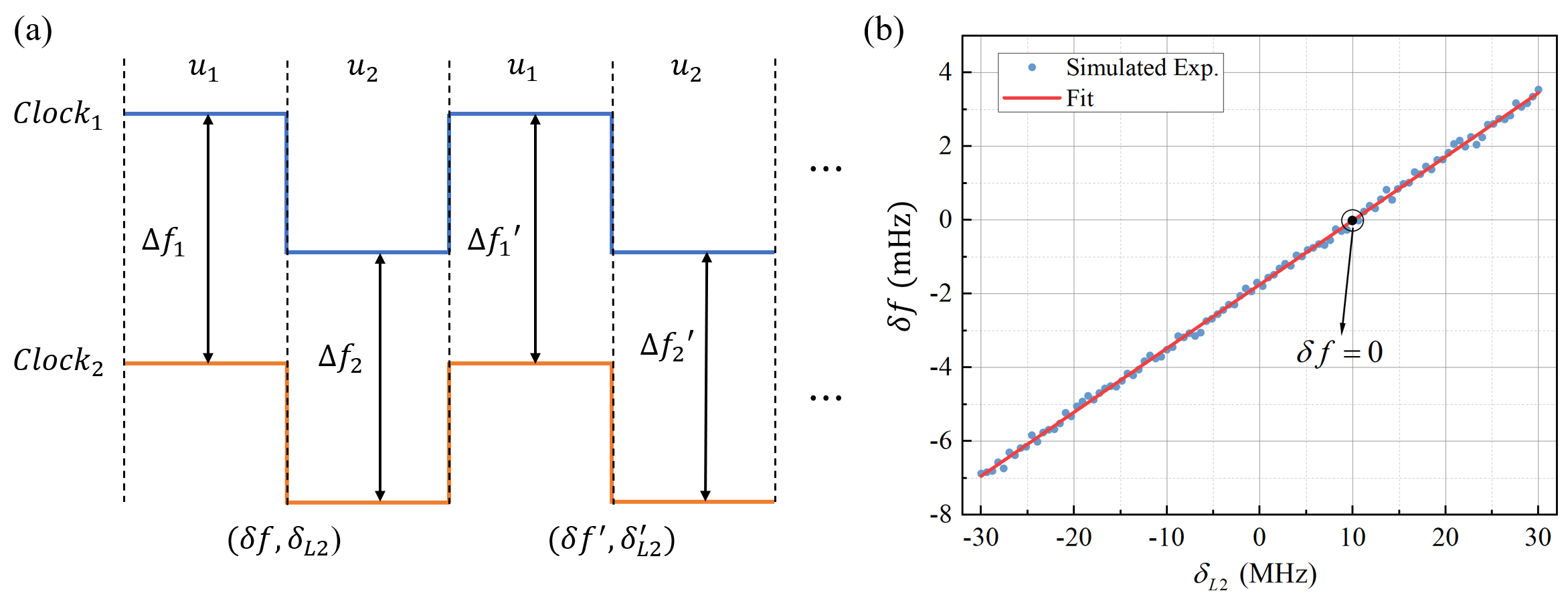}
    \caption{(a) Timing sequence for synchronous differential measurements. (b) Experimental values from simulation and their corresponding fitting curve.}
\end{figure}

\subsubsection{density shift}
Achieving cancellation of the density shift relies on balancing the contributions from on-site \textit{p}-wave and off-site \textit{s}-wave interactions. These interactions depend on the spatial overlap of the atomic wavefunctions, which, in a tilted optical lattice under gravity, are described by \textit{Wannier-Stark states}. The wavefunction centered at site $n$ is given by:
\begin{equation}
W_n(Z) = \sum_{m} \mathcal{J}_{m-n} \left( \frac{2J_0}{Mg a_L} \right) w(Z - ma_L),
\end{equation}
where $\mathcal{J}_n(x)$ are Bessel functions of the first kind, $w(Z)$ is the Wannier function centered at $Z=0$, $a_L = \lambda_L/2$ is the lattice spacing, $g$ is the gravitational acceleration, and $J_0$ is the nearest-neighbor tunneling energy in the lowest band. The key dimensionless parameter governing the spatial extent of the Wannier-Stark states is:
\begin{equation}
\frac{2J_0}{Mg a_L \cos\theta},
\end{equation}
where $\theta$ is the angle between the lattice axis and the direction of gravity.

We derive the angle-dependent magic trap depth $V_0^{\text{magic}}(\theta)$ by first quantifying how the tunneling energy $J_0$ depends on the lattice depth $V_0$. The tunneling energy is approximated by:
\begin{equation}
J_0 \approx \left( \frac{4}{\sqrt{\pi}} \right) E_{\text{rec}}^{1/4} V_0^{3/4} \exp\left[-2\sqrt{\frac{V_0}{E_{\text{rec}}}} \right].
\end{equation}

We evaluate this expression numerically by stepping through values of $V_0$ in the shallow-lattice regime and computing the corresponding $J_0$. Using interpolation, we then invert this relationship to determine $V_0(J_0)$.

\begin{figure}[h]
    \centering
    \includegraphics[width=14cm]{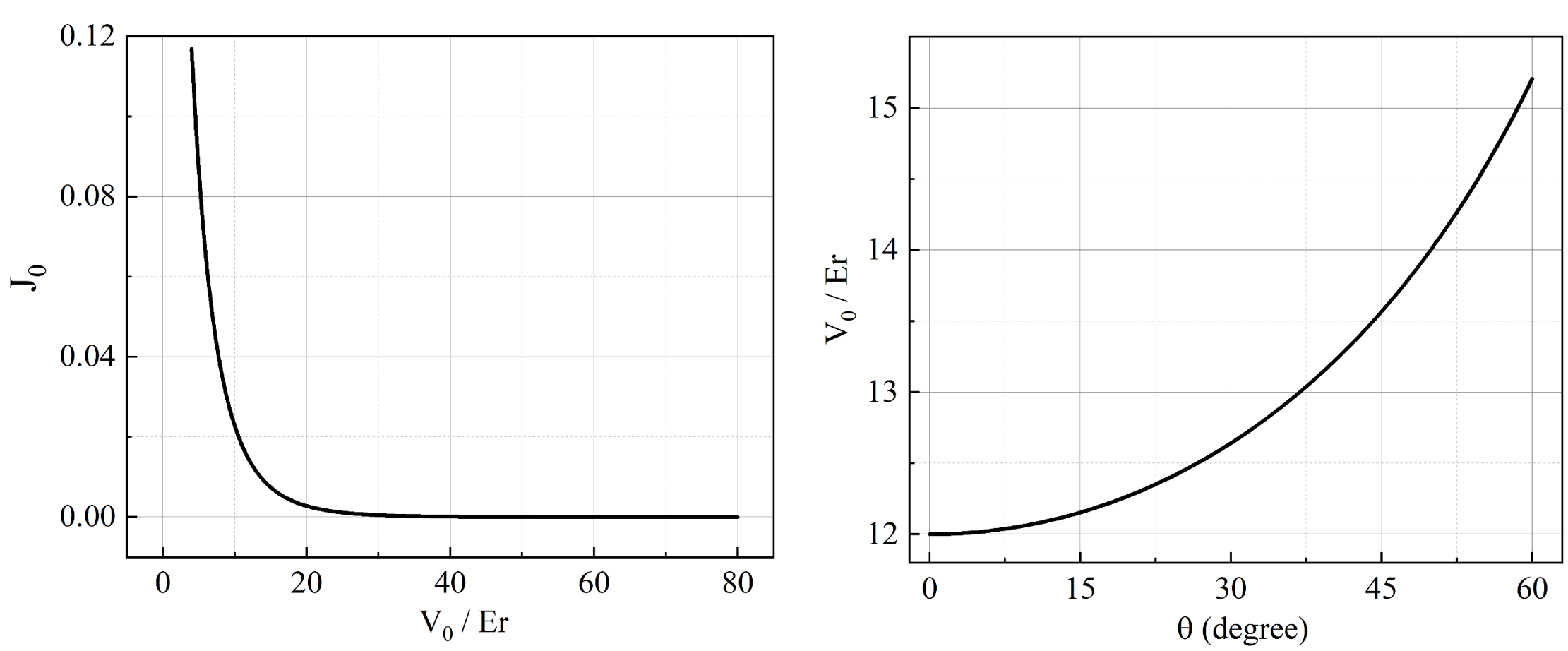}
    \caption{(Left) Tunneling energy as a function of lattice depth. (Right) Magic trapping depth versus tilt angle.}
\end{figure}

According to Ref.~\cite{aeppli2022hamiltonian}, the magic trap depth for a vertically aligned lattice ($\theta = 0$) is $12\,E_{\text{rec}}$, we calculate the value of $\frac{2J_0}{Mg a_L}$ at that point. To preserve the same Wannier-Stark distribution at a finite angle $\theta$, we must keep $\frac{2J_0}{Mg a_L \cos\theta}$ constant. This condition leads to a modified tunneling energy $J_0'$, which we then use to compute the required $V_0^{\text{magic}}(\theta)$ through the previously obtained $V_0(J_0)$ curve.

For example, at a tilt angle of $\theta = 30^\circ$, we find that the magic trap depth increases modestly from $12\,E_{\text{rec}}$ to approximately $12.8\,E_{\text{rec}}$. This confirms that the cancellation condition can still be propose in a shallow lattice even under a moderate tilt, and thus does not impose unrealistic experimental requirements.

To estimate the residual density shift, we consider a differential measurement approach in which the relative atom numbers in the reference and mesasurement ensembles are deliberately varied while keeping all other experimental parameters fixed. Since the density shift scales nonlinearly with atomic population, this method allows for precise extraction of the shift coefficients. Based on the expected interaction strength, trap geometry, and achievable control over atom number ratios, we estimate that this technique enables characterization and subtraction of residual density shifts at a level below $1 \times 10^{-19}$.

\subsection{least-squares fit and Monte Carlo sample}
To mitigate the uncertainty associated with the dynamic BBR shift, we optimize the selected set of Einstein $A$ coefficients by leveraging available experimental data. Our approach aims to estimate the relevant $A$ coefficients in a self-consistent manner, utilizing multiple experimental observables. We model the key Einstein $A$ coefficients as Gaussian variables and compute the experimental observables using the sum-over-states polarizability formula:
\begin{equation}
\alpha_i(\omega) = \alpha_i^{\text{core}} + 2\pi \epsilon_{0} c^3 \sum_k \frac{(2J_k + 1)}{(2J_i + 1)} \frac{A_{ki}}{\omega_{ki}^2 (\omega_{ki}^2 - \omega^2)},
\end{equation}
where $\alpha_i^{\text{core}}$ represents the core-electron polarizability for state $i$, $A_{ki}$ denotes the Einstein coefficient for the transition from state $k$ to state $i$, $\omega_{ki}$ is the transition frequency, and $J_k$ and $J_i$ are the angular momenta of states $k$ and $i$, respectively. The summation encompasses nearly all intermediate states. Using this formula, we calculate various observable quantities $O^{\text{calc}}$ that depend on polarizability, including the differential static polarizability, the 813 nm magic wavelength, the 390 nm magic wavelength, and the tune-out wavelength. The experimental values employed in these calculations are detailed in Table~2.

\begin{table}[h]
      \centering
      \caption{Experimental values selected for fitting the Einstein coefficients.}
      \begin{ruledtabular}
      \begin{tabular}{l c l}
         
          $O^{\text{calc}}$ & Value & Reference \\
          \hline
          $\alpha((5s^2)\, ^1S_0, \nu = 0)$ & $3.07(24) \times 10^{-39} \text{ Cm}^2\text{V}^{-1}$ & \cite{schwartz1974measurement} \\
          $\Delta\alpha(\nu = 0)$ & $4.07873(11) \times 10^{-39} \text{ Cm}^2\text{V}^{-1}$ & \cite{middelmann2012high} \\
          Magic wavelength near 813 nm, $\nu_{813}^{\text{magic}}$ & $368\,554\,825.9(4) \text{ MHz}$ & \cite{kim2023evaluation} \\
          $\partial \Delta\alpha / \partial\nu(\nu = \nu_{813}^{\text{magic}})$ & $1.859(5) \times 10^{-11}$ & \cite{kim2023evaluation}\\
          Magic wavelength near 390 nm, $\nu_{390}^{\text{magic}}$ & $768\,917(18) \text{ GHz}$ & \cite{takamoto2009prospects}\\
          Tune-out wavelength near 689 nm, $\nu_{\text{to}}$ & $434\,972\,130(10) \text{ MHz}$ & \cite{heinz2020state}\\
          $\alpha(^3P_0, \nu = \nu_{\text{to}})$ & $2.564(13) \times 10^{-38} \text{ Cm}^2\text{V}^{-1}$ & \cite{heinz2020state}\\
         
      \end{tabular}
\end{ruledtabular}
      \label{tab:magic_wavelengths}
\end{table}
  
To implement this method, we minimize the following objective function:
\begin{equation}
\chi^2 = \sum_j \left( \frac{O_j^{\text{calc}} - O_j^{\text{exp}}}{\sigma_j} \right)^2,
\end{equation}
where \( O_j^{\text{exp}} \) denotes the experimental value of the observable, \( \sigma_j \) is its associated uncertainty, and \( O_j^{\text{calc}} \) is the corresponding calculated value. When the Einstein A coefficients are varied within their uncertainty ranges, \( O_j^{\text{calc}} \) will change, and the corresponding \( \chi^2 \) will also vary. 
By minimizing \( \chi^2 \), we determine the optimal set of \( A \) coefficients that best reproduce the experimental observables.
As shown in Table 4, we selected six, seven, and eight Einstein coefficients to construct the 6A, 7A, and 8A fitting models, respectively.

Based on the 7A model, we employ four distinct optimization methods to minimize \( \chi^2 \): Differential Evolution (DE), Simulated Annealing (SA), the gradient-based Minimize algorithm from SciPy, and Least Squares (LS). Each method offers unique advantages and limitations:

\begin{enumerate}
  \item \textbf{Differential Evolution (DE)}: A global optimization algorithm particularly effective for high-dimensional, non-convex, and multi-modal problems. Despite its robustness, it is computationally expensive and exhibits slow convergence.
  \item \textbf{Simulated Annealing (SA)}: Inspired by the physical annealing process, this method is capable of escaping local minima and is well-suited for combinatorial optimization. On the other hand, its convergence rate is highly dependent on the choice of temperature parameters.
  \item \textbf{Minimize}: A gradient-based optimizer that performs efficiently for problems with well-defined gradients and is effective for convex or properly initialized non-convex problems. Nevertheless, it may converge to local minima in complex landscapes.
  \item \textbf{Least Squares (LS)}: A widely used method for data fitting, offering fast computation and effective minimization of squared residuals. However, it is sensitive to outliers and restricted to optimizing squared loss functions.
\end{enumerate}

\begin{table}[h]
      \centering
      \caption{Dynamic blackbody radiation shifts calculated using different optimization algorithms.}
      \begin{ruledtabular}
      \begin{tabular}{lcll}

      Method                   & Dynamic BBR Shift (mHz) & Residual & Time    \\\hline
      Differeitial Evolution 1 & -153.154                & 71938.48 & 16min   \\
      Differeitial Evolution 2 & -153.122                & 71958.33 & 16min   \\
      Differeitial Evolution 3 & -153.127                & 71996.72 & 16min   \\
      Simulated Annealing      & -152.75                 & 74057.64 & 29min   \\
      Minimize                 & -152.761                & 73975.12 & 1min40s \\
      Least Square             & -153.131                & 71893.34 & 23s     \\
      J.Y\cite{aeppli2024clock}           & -153.06(32)             &          &         \\ 
      \end{tabular}
\end{ruledtabular}
      \label{shift}
\end{table}

Given the high-dimensional nature of our residual function (seven parameters), a global search for the optimal solution is essential. Additionally, for Monte Carlo sampling calculations, each optimization run must converge as efficiently as possible. As summarized in Table~3, we compare the convergence behavior and optimization results of each method. The results from DE and SA suggest that the global minimum of the residual function is approximately \(-153\) mHz. While Minimize and LS converge to local optima, their results are close to the global minimum identified by DE and SA, making them viable alternatives for minimizing the residual function. Since LS achieves the lowest residual and the shortest computation time per run, we adopt LS as our optimization method and perform Monte Carlo sampling based on it. Based on the LS method, the optimized Einstein A coefficients for the 6A, 7A, and 8A models are presented in Table 4.

In the table, we categorize the transitions requiring optimization into two groups, R and F. For transitions of type R, due to their significant impact on the final results, the coefficient \(A\) is constrained to vary strictly within its uncertainty range during optimization. In contrast, for transitions of type F, which have a relatively minor influence on the final results and residuals, the coefficient \(A\) is treated as a free parameter and allowed to vary over a broader range around the recommended values provided in the literature. The computational results in Table 4 show that, for the 7A model, two out of the four R-type parameters are at the constraint boundaries after optimization, indicating that this model does not adequately describe the experimental data. For the 6A model, only one R-type parameter is at the constraint boundary, while for the 8A model, all R-type parameters remain within the constraints after optimization. Therefore, in calculating the dynamic BBR shift, we only used the optimization results from the 6A and 8A models.

\begin{table}[]
\caption{Optimized Einstein coefficients obtained using the least-squares fitting method. Type R denotes transitions where the coefficient A is constrained to vary within the range of its uncertainty during fitting, while Type F refers to transitions where the coefficient A is allowed to vary freely during fitting.}
\begin{ruledtabular}
\begin{tabular}{lllccc}
\multicolumn{1}{c}{Type} & \multicolumn{1}{c}{Transition}                     & $\lambda$                & Optimized A (s$^{-1}$)   & A Coefficient (s$^{-1}$)                  & Ref.                     \\ \hline
\multirow{3}{*}{R}       & \multirow{3}{*}{$(5s5p)\,^1P_1 \to (5s^2)\,^1S_0$} & \multirow{3}{*}{461 nm}  & $1.8998 \times 10^8$(6A) & \multirow{3}{*}{$1.9001(14) \times 10^8$} & \multirow{3}{*}{\cite{yasuda2006photoassociation}} \\
                         &                                                    &                          & $1.9007 \times 10^8$(7A) &                                           &                          \\
                         &                                                    &                          & $1.8999 \times 10^8$(8A) &                                           &                          \\ \cline{2-5}
\multirow{3}{*}{R}       & \multirow{3}{*}{$(5s5p)\,^3P_1 \to (5s^2)\,^1S_0$} & \multirow{3}{*}{689 nm}  & $46951.5$(6A)            & \multirow{3}{*}{$46888(68)$}              & \multirow{3}{*}{\cite{nicholson2015systematic}} \\
                         &                                                    &                          & $46956$(7A)              &                                           &                          \\
                         &                                                    &                          & $46955.4$(8A)            &                                           &                          \\ \cline{2-5}
\multirow{3}{*}{F}       & \multirow{3}{*}{$Rydberg \to (5s^2)\,^1S_0$}       & \multirow{3}{*}{218 nm}  & $7.034 \times 10^8$(6A)  & \multirow{3}{*}{$1.65 \times 10^8$}       & \multirow{3}{*}{\cite{lisdat2021blackbody}} \\
                         &                                                    &                          & $9.99 \times 10^7$(7A)   &                                           &                          \\
                         &                                                    &                          & $2.47 \times 10^8$(8A)   &                                           &                          \\ \cline{2-5}
\multirow{3}{*}{R}       & \multirow{3}{*}{$(5s5s)\,^3S_1 \to (5s5p)\,^3P_0$} & \multirow{3}{*}{679 nm}  & $8.414 \times 10^6$(6A)  & \multirow{3}{*}{$8.348(66) \times 10^6$}  & \multirow{3}{*}{\cite{garcia1988transition}} \\
                         &                                                    &                          & $8.414 \times 10^6$(7A)  &                                           &                          \\
                         &                                                    &                          & $8.383 \times 10^6$(8A)  &                                           &                          \\ \cline{2-5}
\multirow{3}{*}{R}       & \multirow{3}{*}{$(5s4d)\,^3D_1 \to (5s5p)\,^3P_0$} & \multirow{3}{*}{2603 nm} & $2.76126\times 10^5$(6A) & \multirow{3}{*}{$2.7619(64) \times 10^5$} & \multirow{3}{*}{\cite{aeppli2024clock}} \\
                         &                                                    &                          & $2.7613\times 10^5$(7A)  &                                           &                          \\
                         &                                                    &                          & $2.76131\times 10^5$(8A) &                                           &                          \\ \cline{2-5}
R                        & $(5s5d)\,^3D_1 \to (5s5p)\,^3P_0$                  & 483 nm                   & $3.325\times 10^7$(8A)   & $3.3(2) \times 10^7$                      & \cite{sansonetti2010wavelengths}                  \\ \cline{2-5}
\multirow{3}{*}{F}       & \multirow{3}{*}{$(5s6d)\,^3D_1 \to (5s5p)\,^3P_0$} & \multirow{3}{*}{394 nm}  & $6.383 \times 10^7$(6A)  & \multirow{3}{*}{$1.43 \times 10^7$}       & \multirow{3}{*}{\cite{porsev2008determination}} \\
                         &                                                    &                          & $7.1 \times 10^7$(7A)    &                                           &                          \\
                         &                                                    &                          & $5.8 \times 10^7$(8A)    &                                           &                          \\ \cline{2-5}
\multirow{2}{*}{F}       & \multirow{2}{*}{$Rydberg \to (5s5p)\,^3P_0$}       & \multirow{2}{*}{361 nm}  & $1.003 \times 10^7$(7A)  & \multirow{2}{*}{$4.2 \times 10^7$}        & \multirow{2}{*}{\cite{von1969doppelresonanzuntersuchung}} \\
                         &                                                    &                          & $1.007 \times 10^7$(8A)  &                                           &                         
\end{tabular}
\end{ruledtabular}
\label{tab:einstein_coefficients}
\end{table}

We employed the Leave-One-Out (LOO) method for the 6A, 7A, and 8A models to comprehensively evaluate the influence of each observable on the final dynamic BBR shift results. As detailed in Table~5, we iteratively excluded one observable at a time, performed the fitting procedure, and analyzed the resulting outcomes statistically. The smallest residual was achieved when the 813 nm magic wavelength was omitted from the fitting process. Based on this observation, we defined the ratio of the residual after removal to the residual from the full dataset fitting as the inflation factor for the final uncertainty estimation. Based on the residuals listed in Table 5, the derived inflation factors for the 6A, 7A, and 8A models are 2.07, 4.27, and 2.2, respectively.

\begin{table}[]
\caption{Dynamic BBR shifts and corresponding residuals obtained using the leave-one-out method.}
\begin{ruledtabular}
\begin{tabular}{lccc}

\multicolumn{1}{l}{Item}                                     & Fit model & Freq(mHz) & Sqrt Residual \\ \hline
\multirow{3}{*}{All}                                         & 6A        & -153.03   & 72460         \\
                                                             & 7A        & 153.13    & 71893         \\
                                                             & 8A        & -152.77   & 74099         \\ \hline
\multirow{3}{*}{Leave 813nm Magic wavelength out}            & 6A        & -152.36   & 50325         \\
                                                             & 7A        & -151.26   & 35406         \\
                                                             & 8A        & -153.08   & 49957         \\ \hline
\multirow{3}{*}{Leave 390nm Magic wavelength out}            & 6A        & -153.25   & 64202         \\
                                                             & 7A        & 154.25    & 65809         \\
                                                             & 8A        & -152.85   & 63710         \\ \hline
\multirow{3}{*}{Leave $\Delta\alpha(\nu = 0)$}               & 6A        & -152.86   & 60815         \\
                                                             & 7A        & 154.14    & 66755         \\
                                                             & 8A        & 152.72    & 67186         \\ \hline
\multirow{3}{*}{Leave $\partial \Delta\alpha / \partial\nu$} & 6A        & -153.09   & 61792         \\
                                                             & 7A        & -154.09   & 70695         \\
                                                             & 8A        & -152.77   & 64098         \\ 
\end{tabular}
\end{ruledtabular}

\end{table}

To comprehensively evaluate the uncertainty of the fitting results, we employed Monte Carlo sampling for the 6A, 7A, and 8A fitting models. Random sets of experimental observables \(\{ x_i \}\) were drawn from their respective distributions (e.g., \( x_i \sim \mathcal{N}(\mu, \sigma) \), where \(\mathcal{N}(\mu, \sigma)\) denotes a Gaussian distribution with mean \(\mu\) and standard deviation \(\sigma\)). Since the least-squares fitting method used in our analysis is a nonlinear optimization, even minor variations in the experimental data sets within their uncertainty ranges could lead to dramatically different optimization paths and substantially altered results compared to the unperturbed case. To obtain a statistically meaningful Monte Carlo sampling distribution, for each set of observables, our code ensures that the residual function follows a fixed optimization path to obtain the optimal set of Einstein A coefficients and compute the dynamic BBR shift. We performed 20,000 Monte Carlo samplings for the 6A, 7A, and 8A fitting models and statistically analyzed the results, including Gaussian fitting. The complete Monte Carlo datasets for each model are available in \cite{jin2025sr}.

\begin{figure}[htbp]
      \centering
      \includegraphics[width = 16cm]{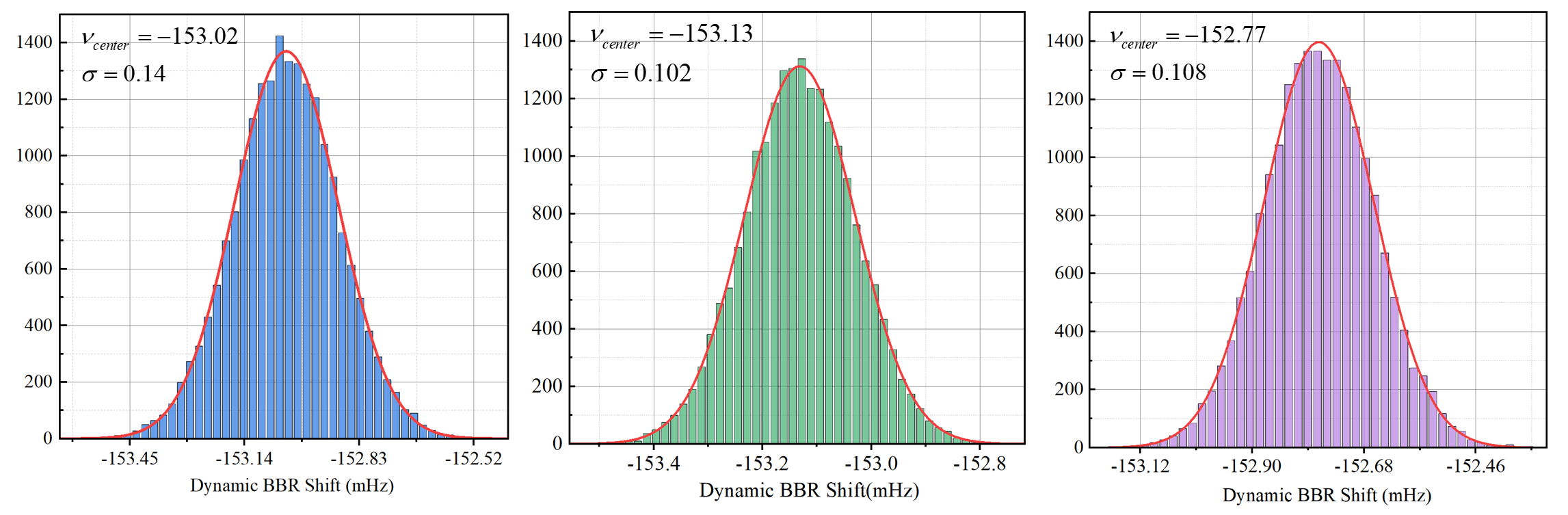}
      \caption{Distribution of Monte Carlo sampling results for the dynamic BBR shift at 300 K, showing results from the 6A (left), 7A (middle), and 8A (right) fitting models.}

\end{figure}

Figure 8 presents the statistical distributions from Monte Carlo sampling of different optimization models, with red curves showing Gaussian fits to the results. All three models exhibit good agreement between the fits and statistical distributions. The central values from Gaussian fits yield dynamic BBR shifts of -153.02 mHz (6A), -153.13 mHz (7A), and -152.77 mHz (8A), with standard deviations $\sigma$ of 0.14 mHz, 0.102 mHz, and 0.108 mHz, respectively. After applying inflation factors determined via the leave-one-out method, the final uncertainties become 0.29 mHz (6A), 0.43 mHz (7A), and 0.23 mHz (8A). The 7A model fails to adequately describe the experimental data (with two R-type parameters reaching constraint boundaries after optimization), leading us to exclude it from our final determination. By combining results from the robust 6A and 8A models, we obtain a final dynamic BBR shift of -152.89(29) mHz at 300 K. Figure 9 compares these results with previous calculations, demonstrating that our 6A+8A combined value agrees with the JILA Collaboration's (2024) result within uncertainties\cite{aeppli2024clock}.

\begin{figure}[htbp]
      \centering
      \includegraphics[width = 6cm]{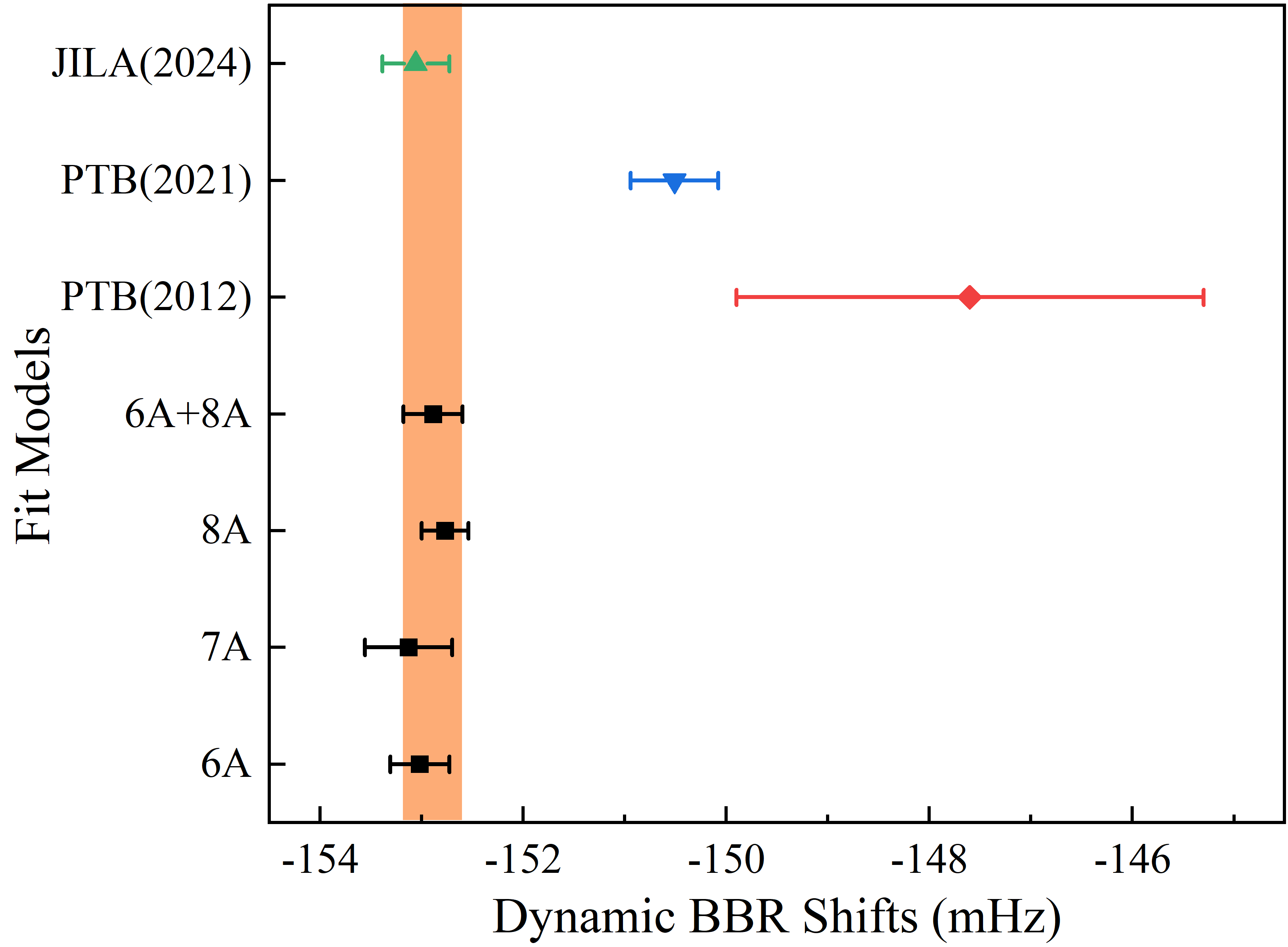}
      \caption{Calculated dynamic BBR shifts for different optimization models compared with previous results. The final recommended value is derived from the combined 6A+8A model results.}

\end{figure}

We calculated the blackbody radiation shift from 10 K to 400 K in 1 mK increments. For temperature measurements, we will determine the corresponding frequency shift at any given temperature using a lookup table with linear interpolation. We simultaneously computed three datasets: the mean blackbody radiation shift, upper uncertainty bound, and lower uncertainty bound. As shown in Figure 1 of the main text, these calculations enabled us to determine the temperature measurement errors caused by blackbody radiation uncertainty at different temperatures, as presented in Figure 4 (a).

\subsection{Atomic Cloud Surface Radiation Analysis}
To quantitatively evaluate our blackbody radiation shield's effectiveness, we computed the environmental radiation intensity distribution at the atomic cloud surface using surface-to-surface radiation finite-element simulations with the hemicube method (radiation resolution: 1024). We analyzed two distinct configurations: external to the chamber at 5 mm from the cavity and at the chamber center, as shown in Fig. 10.

\begin{figure}[htbp]
      \centering
      \includegraphics[width = 9cm]{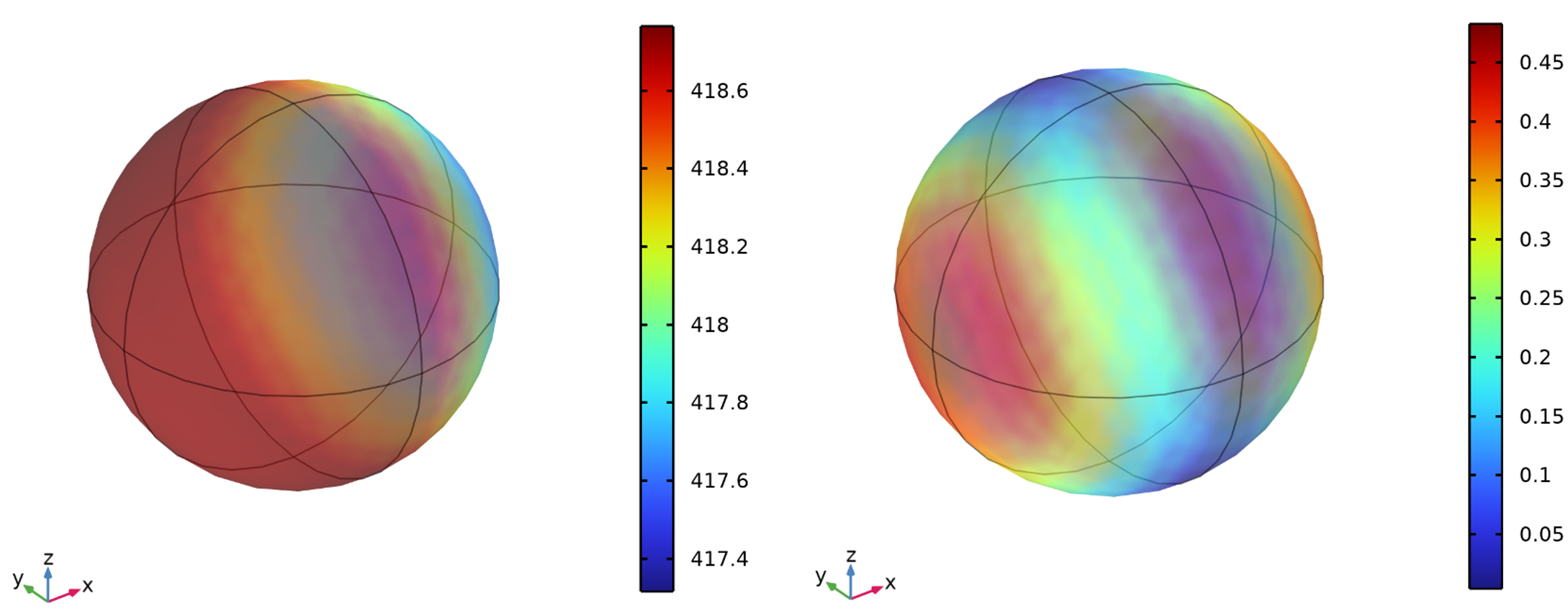}
      \caption{(Left) Environmental radiation distribution at the atomic cloud surface 5 mm outside the shielding cavity. (Right) Environmental radiation distribution at the atomic cloud surface inside the shielding cavity.}

\end{figure}

Integrating the radiation intensity distributions yields total power values of $3.006\times10^{-4}$ W (external) and $1.66\times10^{-7}$ W (chamber center). The chamber thus blocks $3.004\times10^{-4}$ W of environmental radiation, corresponding to a 99.88\% shielding efficiency. In our design, both the reference and measurement cavity walls are coated with carbon nanotube films. As discussed in the main text, the forest-like microstructure of these carbon nanotubes introduces an angular dependence in their emissivity. We model this angular dependence using $\varepsilon_{\text{eff}} = \varepsilon_0(1 - R(\theta))$, where $R(\theta)$ is the angle-dependent reflectance. Fitting experimental data from Ref.~\cite{mizuno2009black} for single-walled carbon nanotube (SWCNT) coatings yields $R(\theta) = \beta\sin^2\theta$, with $\beta = 0.03$ and $\varepsilon_0 > 0.98$. For conservative uncertainty estimation, we adopt enhanced parameters: $\beta = 0.1$ and $\varepsilon_0 = 0.98$. Our numerical approach discretizes the atomic channel and cloud into thousands of surface elements. For each cloud element, we compute the effective emissivity through numerical integration of contributions from all visible cavity wall segments, as illustrated in Fig.~11.

\begin{figure}[htbp]
      \centering
      \includegraphics[width = 6cm]{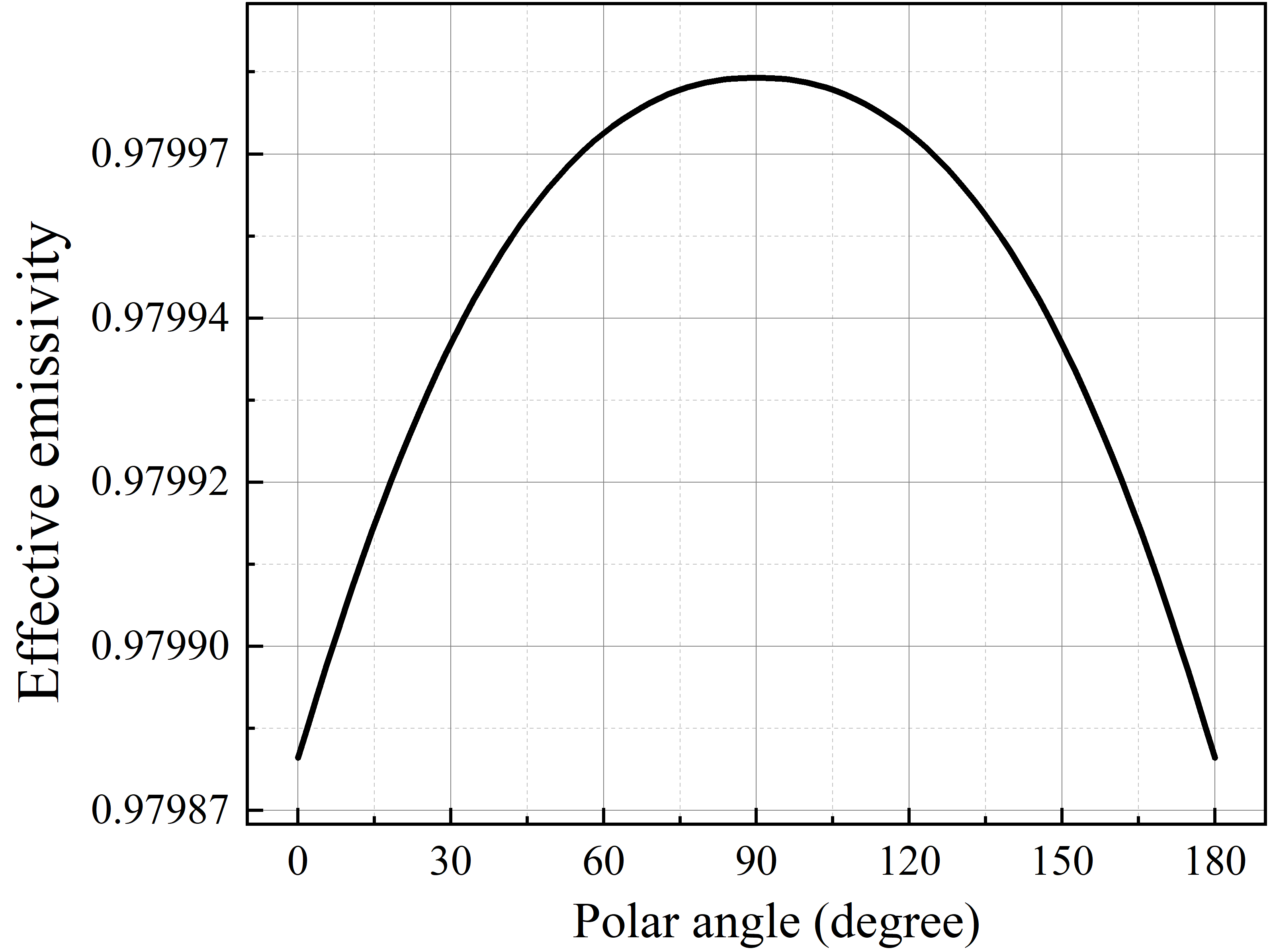}
      \caption{Effective emissivity of blackbody radiation incident on the atomic cloud surface from the atomic channel as a function of polar angle (polar axis parallel to the channel)}

\end{figure}

Figure~11 shows the effective emissivity variation remains minimal across all polar angles. This behavior arises from two key factors: first, the carbon nanotube forest structure maintains near-blackbody emissivity even at grazing angles; second, our long cavity design ensures most atomic radiation originates from distant walls at near-normal incidence.

We use the statistically averaged emissivity for calculations, with the inflated full variation range (0.005) treated as uncertainty. Propagating this through our model, the effective blackbody temperature experienced by atoms is:
\begin{equation}
T_{eff}^4 = \frac{{\Omega T_2^4 + (4\pi  - \Omega )\varepsilon T_1^4}}{{\Omega  + (4\pi  - \Omega )\varepsilon }},
\end{equation}
where $\Omega = 0.0069$ is the calculated environmental radiation solid angle, and $T_1$ and $T_2$ represent the measurement cavity and environmental temperatures, respectively. 

\begin{figure}[htbp]
      \centering
      \includegraphics[width = 6cm]{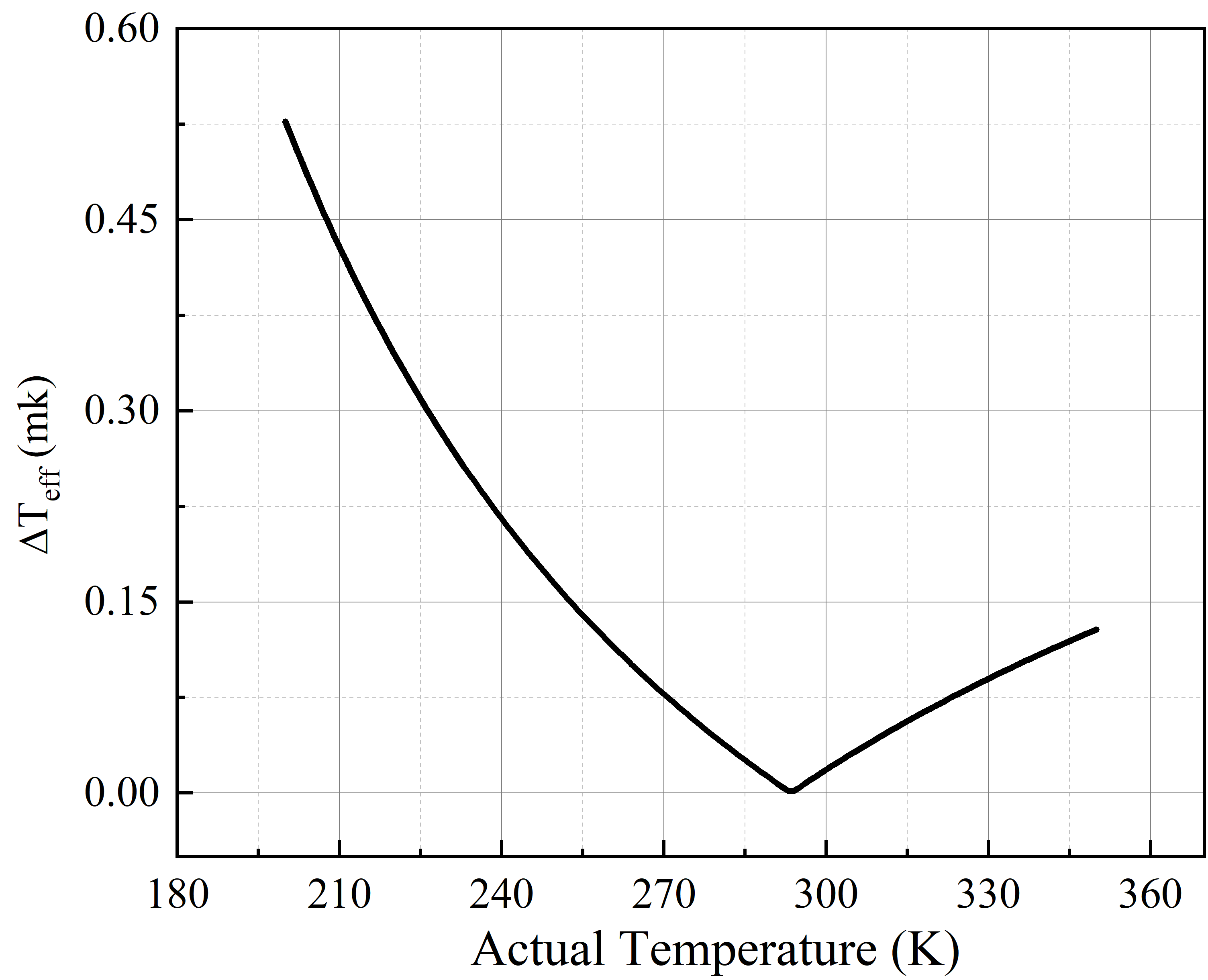}
      \caption{Variation of the effective blackbody radiation temperature at the atomic channel surface as a function of the actual temperature}

\end{figure}

As evident from the Fig.~12, within the temperature calibration range of 200 K to 350 K, the temperature calibration uncertainty induced by emissivity variations remains below 0.6 mK. Consequently, we have incorporated this 0.6 mK contribution into the overall calibration budget as a systematic error component.

Using the calculated effective emissivity of the atomic channel surfaces, we determined the radiation distribution at the atomic ensemble surfaces in both reference and measurement chambers, as shown in Fig.~3.

We further examined the effect of environmental radiation asymmetry between chambers by modeling temperature gradients in the outer vacuum cavity. Following Ref.~\cite{hobson2020strontium}, where distributed temperature sensors measured peak-to-peak variations of 350 mK across the vacuum cavity surface, we analyzed the effect of such thermal gradients. Figure~13 shows that this 350 mK gradient induces a maximum radiation flux variation of just $2.29\times10^{-3}$ W/m$^2$ at the atomic ensemble surface -- a negligible contribution. We further calculated that a 350 mK temperature difference between the reference and measurement chambers' environments would produce only a $1.22\times10^{-20}$ fractional frequency shift in differential measurements. These analyses confirm the design's robustness against environmental thermal inhomogeneities.
\begin{figure}[htbp]
      \centering
      \includegraphics[width = 6cm]{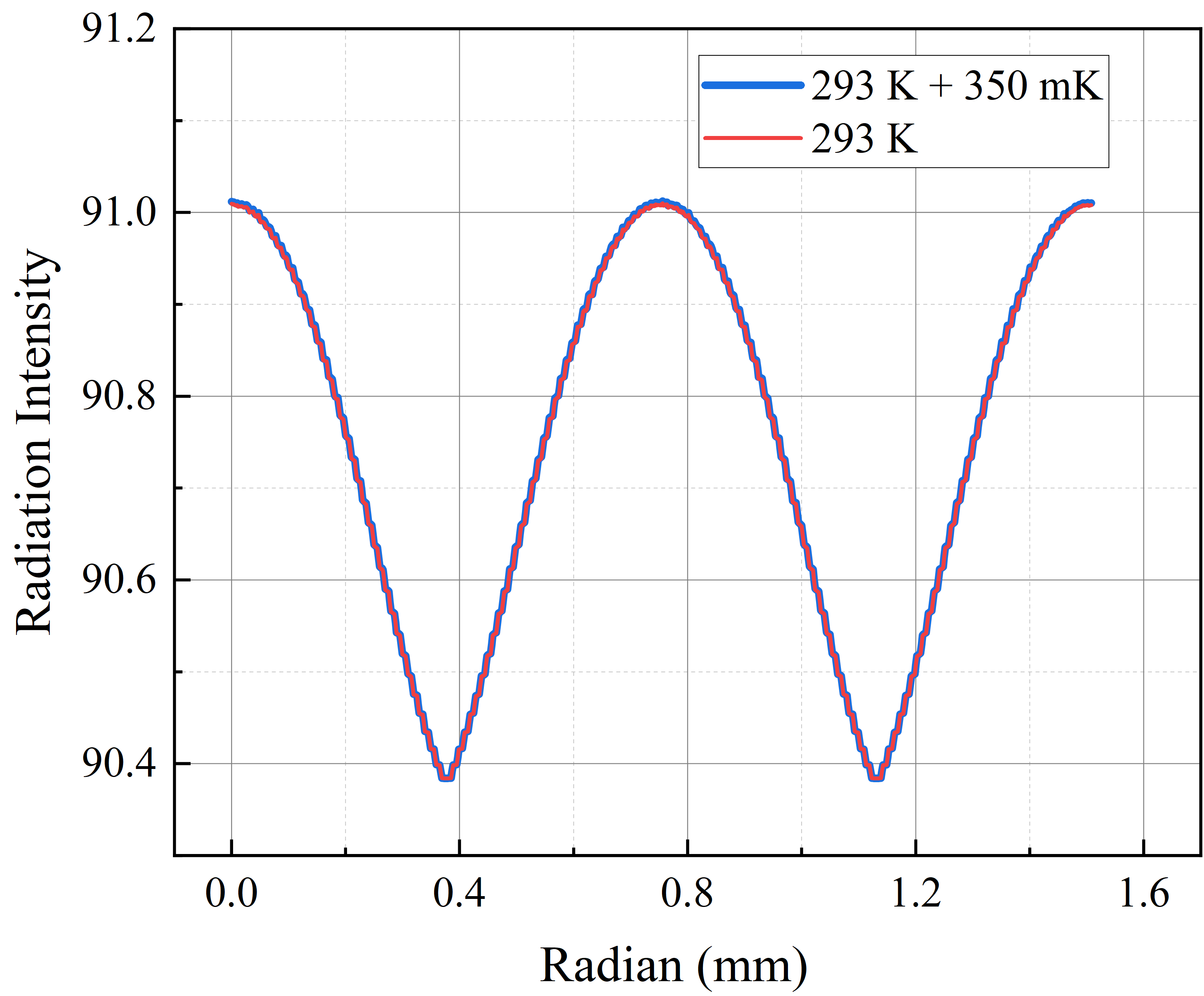}
      \caption{Radiative intensity distributions across atomic cloud surfaces for measurement chambers with  350 mK environmental temperature differences.}

\end{figure}

\subsection{Temperature Gradients in Measurement and Reference Chambers}
We performed a time-domain heat transfer finite element multiphysics simulation for both the measurement chamber and the reference chamber. For the reference chamber simulation, we constructed the geometric model shown in Fig.~2, which consists of an upper blackbody radiation shielding chamber and a lower heat-conducting copper block.

Before running the simulation, we assigned material properties to the geometric model. The physical properties of the materials used in the model are summarized in Table~6.

Next, we applied comprehensive physical boundary conditions to the model. The reference and measurement chambers in our apparatus feature all-copper construction with polished surfaces ($\varepsilon \ll 1$). Three key design elements suppress thermal perturbations: (i) high surface reflectivity minimize thermal radiation absorption from the vacuum enclosure, (ii) future implementations will incorporate thermal isolation between the cryocooler cold head and vacuum chamber, and (iii) active stabilization maintains the reference cavity temperature via real-time feedback control of the pulse tube cryocooler power. This comprehensive approach suppresses vacuum-enclosure-induced thermal gradients, maintaining both radiative and conductive heat transfer components at negligible levels during operation. The atomic channel surface is coated with a high-emissivity material, and the divergence of the Gaussian laser beam during propagation through the channel leads to partial absorption at the surface. This establishes a temperature gradient between the channel surface and the platinum resistance thermometer used for temperature measurement. Through careful consideration of the channel surface's absorptivity and effective area, we derived a resultant heat flux of 116.5 W/m². This value was subsequently incorporated as a thermal flux boundary condition in our simulation model.
Thermal contact boundary conditions were imposed at the interface between the cavity and the copper block. The bottom surface of the copper block was assigned a cooling temperature boundary condition, while all other surfaces were treated as thermally insulated.

\begin{table}[h]
      \centering
      \caption{Material Properties for Thermal Conductivity, Specific Heat Capacity, and Density}
      \begin{ruledtabular}
      \begin{tabular}{lccc}
          
          \textbf{Material} & \textbf{Thermal Conductivity} & \textbf{Specific Heat Capacity} & \textbf{Density} \\
          & \textbf{(W/(m·K))} & \textbf{(J/(kg·K))} & \textbf{(kg/m³)} \\
          \hline
          Copper & \( 447.15 - 0.585 T + 0.00226 T^2 \) & \( -215.28 + 8.23 T - 0.047 T^2 \) & \( 9062.24 - 0.39 T \) \\
          Titanium & \( 60.28 - 0.577 T + 0.0043 T^2 \) & \( 229.88 + 1.6 T - 0.0026 T^2 \) & \( 4525.5 - 0.053 T \) \\
          Silicone Grease & 2 & 920 & 2900 \\
          Quartz & \( 1.38 \) & \( -46.3577953 + 2.68428748 T \) & \( 2672.5 - 0.166 T \) \\
          Air & \( -0.00084 + 0.00011 T \) & \( 1011 + 0.044 T \) & \( 352.716 T^{-1} \) \\
          Polystyrene & 0.03 & 1340 & 1060 \\
          Peltier & 2.11 & 165 & 7600 \\
          \hline
      \end{tabular}
      \end{ruledtabular}
\end{table}
      
Currently, commercially available miniature pulse tube cryocoolers can propose a cooling temperature of 4.2 K, although their cooling power at approximately 45 K is limited to about 30 W. To efficiently cool the reference cavity from room temperature (293 K) to approximately 4 K, we propose a staged cooling approach that combines liquid nitrogen evaporative cooling with the pulse tube cryocooler. Initially, liquid nitrogen evaporative cooling is employed to rapidly cool the shielding cavity to about 77 K. Subsequently, the first stage of the pulse tube cryocooler is activated to further reduce the temperature to 45 K. Finally, the second stage of the pulse tube cryocooler lowers the temperature of the shielding cavity to 4.2 K. To accurately simulate the actual cooling process, we incorporated the following piecewise function into the cooling temperature boundary condition:
\begin{equation}
      {T_{cool}} = \left\{ \begin{array}{l}
      77,0 \le t \le 30{\rm{ }}\\
      45 + (77 - 45) \times {e^{(t - 30)/{\tau _1}}},30 < t \le 140\\
      4.2 + (45 - 4.2) \times {e^{(t - 140)/{\tau _2}}},140 < t \le 2000
      \end{array} \right.
\end{equation}

In this equation, \(\tau_1\) and \(\tau_2\) represent the cooling temperature decay factors for the first and second stages of the pulse tube cryocooler, respectively. Here, \(\tau\) is defined as:

\[
\tau = \frac{m C_p T}{P_{\text{cool}}},
\]

where \(m\) is the mass of the model, \(C_p\) is the specific heat capacity of the material, \(T\) is the temperature of the cavity prior to cooling, and \(P_{\text{cool}}\) is the cooling power of the cryocooler. By implementing this piecewise function for the cooling temperature, we calculated the time-dependent temperature variation of the atomic channel surface of the shielding cavity during the cooling process, as illustrated in the figure. The results show that the temperature of the atomic channel surface of the reference cavity can be reduced to 4.2 K after approximately 2000 seconds.

\begin{figure}[htbp]
      \centering
      \includegraphics[width = 9cm]{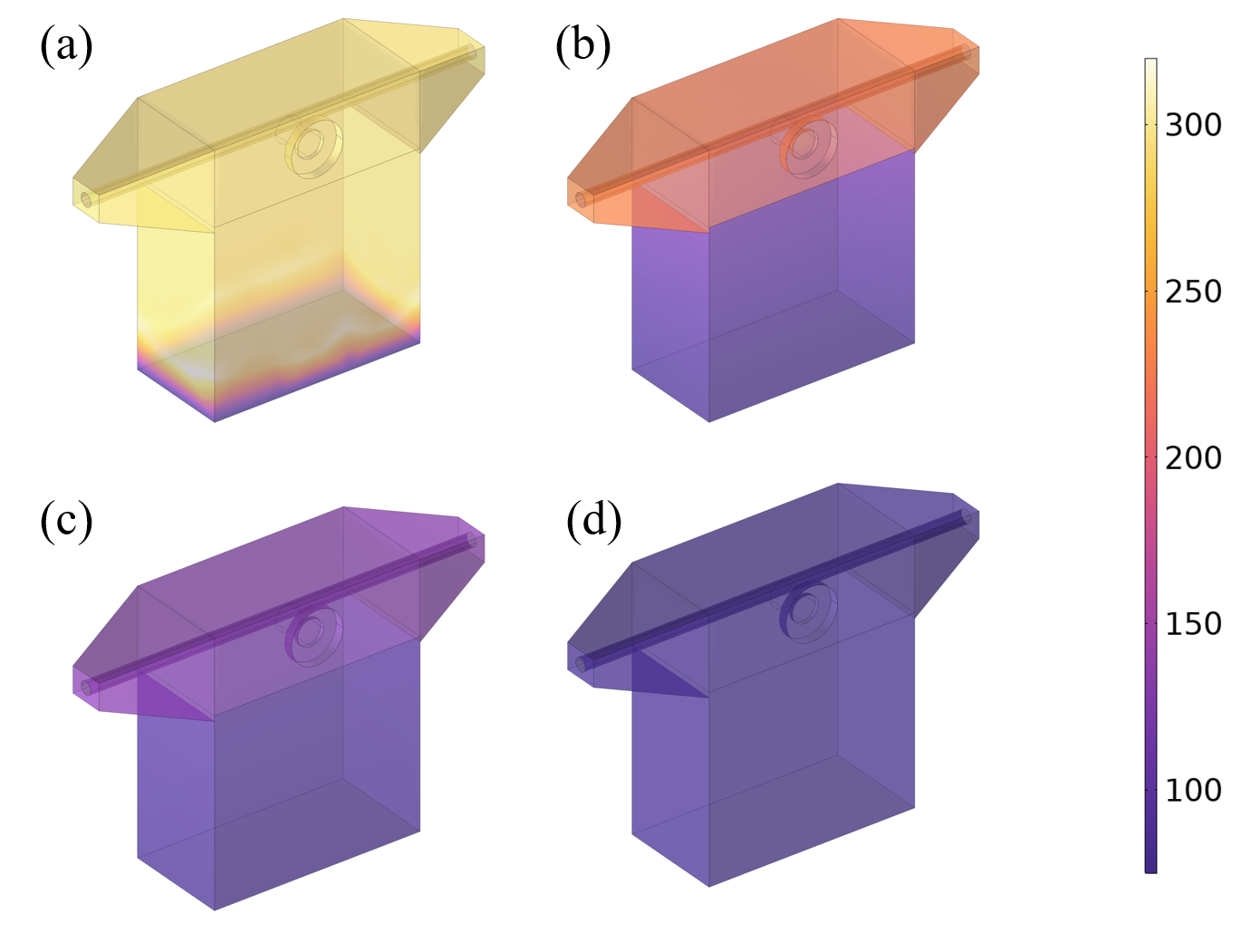}
      \caption{Temperature distribution of the reference chamber at different time points. (a), (b), (c), and (d) correspond to the temperature distributions at 0 s, 3 s, 9 s, and 100 s, respectively.}
      \label{fig7}

\end{figure}

\begin{figure}[h]
      \centering
      \includegraphics[width = 12cm]{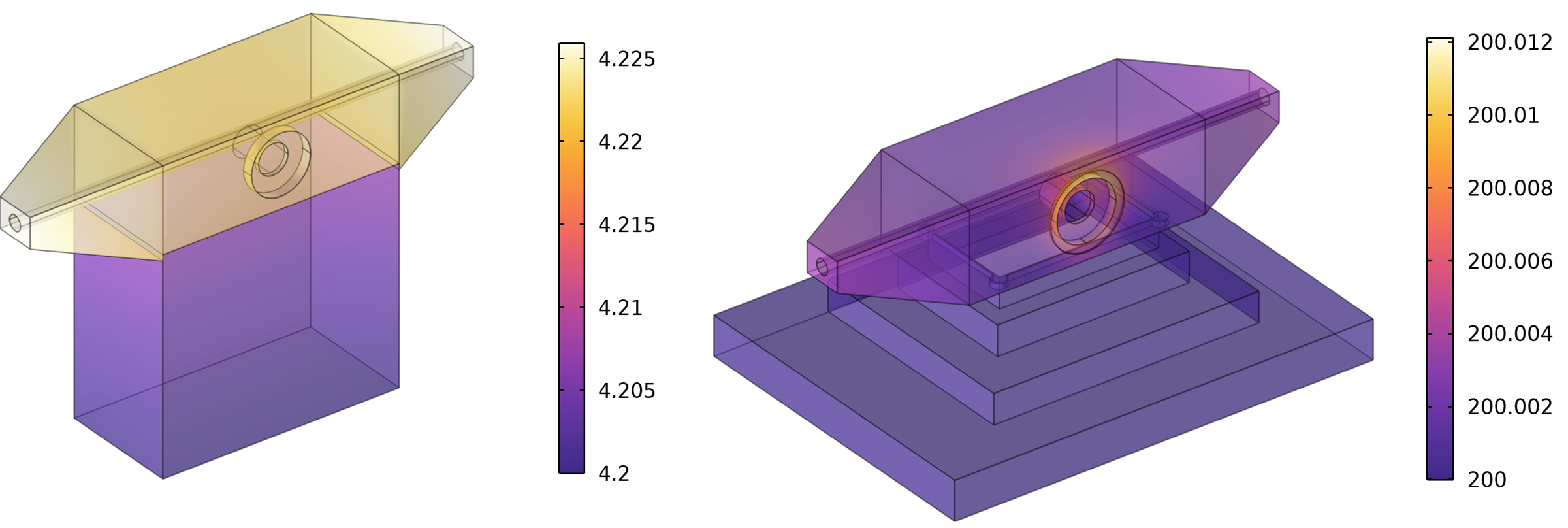}
      \caption{Temperature distribution on the reference chamber(left) and measurement chamber(right) under thermal equilibrium.}
      \label{fig9}

\end{figure}

We also performed time-domain heat transfer finite element simulations for the measurement chamber, with the geometric model illustrated in the figure 2. The structure of the measurement chamber is identical to that of the reference chamber, except that the measurement chamber employs a fourth-order Peltier cooler for temperature control, enabling continuous temperature variation within the range of 200 K to 300 K. A quartz tube, as shown in the figure, connects the scientific chamber to the external environment, facilitating the calibration of the platinum resistance thermometer. To ensure that the temperature of the platinum resistance thermometer closely matches that of the measurement chamber, we filled the surrounding area with silicone grease. Additionally, to minimize the impact of heat convection caused by air flow on the measurement chamber temperature, we inserted a polystyrene plug into the hole in the vacuum chamber wall. The physical properties of the materials added to each component in the finite element simulation are listed in Table~6.

Next, we applied physical boundary conditions to the geometric model of the measurement chamber. In our theoretical design, we implement active temperature stabilization for the measurement chamber through real-time monitoring and thermoelectric voltage regulation. Consequently, we have treated the vacuum-enclosure-induced thermal gradients as negligible for the measurement chamber, applying instead an identical thermal flux boundary condition to that implemented for the atomic channel surfaces in the reference chamber. Thermal contact boundary conditions were assigned to the interfaces between each component, while the surface of the scientific chamber was maintained at a constant temperature of 297.3 K. The interface between the Peltier cooler and the measurement chamber was set as a cooling boundary condition, with the cooling temperature parameterized for scanning over a range of 200 K to 300 K in steps of 5 K. All other external surfaces were treated as thermally insulated boundary conditions. Based on the finite element simulation results, we obtained the temperature gradient, \(\Delta T_3\), between the surface of the atomic channel and the probe as a function of both time and cooling temperature. The computed results are presented in Fig.~4. Furthermore, we calculated the temperature distribution on the reference chamber and measurement chamber under thermal equilibrium., as shown in Fig.~15.

\nocite{*}

\bibliography{apssamp}

\end{document}